\documentclass[manuscript,screen,natbib=false]{acmart}
\usepackage{xurl}
\usepackage[
  datamodel=acmdatamodel,
  style=acmnumeric,
  defernumbers=true,
  swlabels=false,   
]{biblatex}

\DeclareSourcemap{
  \maps[datatype=bibtex]{
    \map{
      \perdatasource{policies.bib}
      \step[fieldset=keywords, fieldvalue=aipolicy, append]
    }
  }
}

\DeclareFieldFormat[software,softwareversion,softwaremodule,codefragment]{swhid}{%
  \mkbibacro{SWHID}\addcolon\addspace
  \ifhyperref
    {\href{https://archive.softwareheritage.org/#1}{\nolinkurl{#1}}}
    {\nolinkurl{#1}}%
}

\DeclareFieldFormat[software,softwareversion,softwaremodule,codefragment]{version}{}

\usepackage{hyperref}

\usepackage{colortbl}

\usepackage{subcaption}
\usepackage{multirow}
\usepackage{booktabs}
\usepackage{color,soul}

\usepackage{tcolorbox}
\usepackage{dblfloatfix}
\usepackage{booktabs}

\newtcolorbox{boxH}{
    colback = white!90!gray, 
    colframe = black, 
    boxrule = 0pt, 
    leftrule = 3pt
}

\newtcolorbox{boxH2}{
    colback = brown!10, 
    colframe = black, 
    boxrule = 0pt, 
    leftrule = 3pt
}

\newcommand{\blue}[1]{\sethlcolor{blue!10}\hl{#1}}
\newcommand{\green}[1]{\sethlcolor{green!10}\hl{#1}}
\newcommand{\red}[1]{\sethlcolor{red!10}\hl{#1}}

\setcopyright{rightsretained}
\copyrightyear{2026}
\acmYear{2026}

\begin{document}

\title{``We Permit the Use of AI, but [...]'': The Landscape of AI Policies in Popular Open Source Projects}

\author{Andre Hora}
\email{andrehora@dcc.ufmg.br}
\orcid{0000-0003-4900-1330}
\affiliation{%
  \department{Department of Computer Science}
  \institution{UFMG}
  \city{Belo Horizonte}
  \country{Brazil}
}

\author{Romain Robbes}
\email{romain.robbes@labri.fr}
\orcid{0000-0003-4569-6868}
\affiliation{%
  \department{INP, LaBRI, UMR 5800}
  \institution{Univ. Bordeaux, CNRS}
  \city{Bordeaux}
  \country{France}
}

\author{Stefano Zacchiroli}
\email{stefano.zacchiroli@telecom-paris.fr}
\orcid{0000-0002-4576-136X}
\affiliation{%
  \department{LTCI}
  \institution{Télécom Paris, Institut Polytechnique de Paris}
  \city{Palaiseau}
  \country{France}
}

\begin{abstract}
Open source communities are converging on a new governance artifact: the \emph{AI contribution policy}, an explicit statement of whether contributors may use generative AI (GenAI), under what conditions, and what they must disclose.
These policies barely existed a few months ago and are being written and adopted now, in public, while the practices they govern are still taking shape.
This provides a rare opportunity to observe a governance convention as it forms rather than reconstruct it afterward.

Public discourse suggests that open source projects are closing their doors to AI-assisted contributions; our data show a different and more nuanced picture.
We analyzed the 2,000 most popular GitHub repositories, complemented by 36 well-known projects and organizations, identified 281 AI contribution policies, and manually classified them along the six dimensions of a purpose-built classification scheme; to study how policies change, we also tracked 92 dedicated AI policy files over time.
We answer four research questions on (1) AI usage allowance, (2) AI disclosure practices, (3) AI slop countermeasures, and (4) AI policy evolution.

We find that, first, permission is the norm rather than the exception: 83.3\% of policies permit or encourage AI in code contributions, and only 14.9\% forbid it.
But permission comes with conditions, as 67.3\% require a high level of human involvement and 43.4\% assign accountability to the human contributor.
Guidance is concentrated on code contributions (98.2\%) and much sparser on communication (42.3\%) and issues, bugs, and security (31.3\%).
Second, AI disclosure is required by 48.8\% of policies, most often in pull request descriptions and commit messages, but what must be disclosed varies widely, and no cross-project convention has emerged.
Third, we identify ten countermeasures against AI slop, targeting pull requests, users, and autonomous agents, the most common being aggressively closing pull requests, banning users, and disallowing fully autonomous agents.
Finally, policies are not static: half of the dedicated AI policy files we track have already been revised since creation, to tighten quality controls, clarify disclosure rules, enforce human accountability, and expand their scope.
Our dataset and classification scheme give maintainers a peer baseline against which to situate their own policy, and give researchers a labeled corpus for studying the impact of AI policies on open source projects.
\end{abstract}


\maketitle

\section{Introduction}

Generative AI (GenAI) is transforming how software is designed, developed, and engineered~\cite{fan2023large, hou2023large}.
Coding agents such as Claude Code and Codex operate with a high degree of \emph{autonomy}: they can invoke external tools, execute code, and complete development tasks end-to-end~\cite{agentminingpaper}.
Their adoption has been fast: a recent large-scale study estimates that, in early 2026, coding agent adoption on GitHub was close to 30\% overall~\cite{robbes2026agentic} and 76\% in new projects~\cite{robbes2026verymuchagentic}.

This shift meets free and open source software (FOSS) at an awkward angle.
Traditional open source development rests on assumptions about who contributes and how: that a patch has a human author who understands it, that authorship can be attributed and vouched for---as artifacts like the Developer Certificate of Origin (DCO)\footnote{\url{https://developercertificate.org/}, accessed 2026-08-28} make explicit---and that reviewer attention is a scarce, often volunteer-funded resource that a contributor's own effort is expected to repay.
Coding agents strain all these assumptions at once, at a volume that current review processes were not designed to absorb~\cite{song2024impact, nakashima2026agentic, yang2026beyond, baltes2026endless}.

The result has been a visible backlash.
Maintainers report an endless stream of AI-generated pull requests and bug reports that are plausible on the surface but worthless on inspection, a phenomenon that has come to be known as \emph{``AI slop''}: low-quality content produced at scale using AI~\cite{ai-slop, baltes2026endless, baltes2026ai}.
The debate has spilled into project governance venues: the LLVM community has argued publicly about how its AI policy interacts with its code of conduct and with everyday practice~\cite{llvm-1, llvm-2}, and the Rust compiler team has moved to empower reviewers to reject burdensome pull requests outright~\cite{rust-lang}.
These are the cases that reach a wide audience, and they have shaped a widely held impression that open source is turning against AI.

Less visible, but far more common, is a quieter response: projects are writing their rules down.
A new kind of artifact has appeared in open source repositories over the past 18 months, the \emph{AI contribution policy}, or \emph{AI policy} for short.
This policy may be documented in a dedicated file, such as \texttt{AI\_POLICY.md}, in a section of \texttt{CONTRIBUTING.md}, or in an agent configuration file, such as \texttt{AGENTS.md}.
The artifact is new: among the 92 dedicated AI policy files we study, only 8 were created in 2025, against 84 in 2026.
In this respect, AI contribution policies follow a familiar trajectory.
Contribution guidelines and codes of conduct also began as ad hoc documents in individual projects before consolidating into conventions with recognizable names, locations, and structures~\cite{tsay2014influence, elazhary2019not, falcucci2025contribution}.
The timing here is singular: this convention is being formed right now, in public; we can observe it while it forms rather than reconstructing it after the fact.

Yet, we still know little about what these policies actually say.
For \emph{contributors}, the expectations attached to AI-assisted work are largely unwritten and vary from project to project, so that following one project's rules may mean violating another's inadvertently.
For \emph{maintainers}, writing a policy currently means guessing: there is no peer baseline describing what comparable projects require, nor evidence about which requirements are common enough that contributors will recognize them.
For \emph{tool and platform builders}, there is no disclosure convention to target; projects disagree even on the mechanics, some requiring a Git commit message trailer such as \texttt{Assisted-by} while others explicitly \emph{forbid} the \texttt{Co-authored-by} trailer that some coding agents emit by default.
For \emph{researchers}, the impact is methodological: mining contributions supported by AI at scale depends on the traces that disclosure rules produce, thus, those rules dictate what can be measured in the first place~\cite{agentminingpaper}.

To characterize these policies systematically, we analyzed the 2,000 most popular GitHub repositories, complemented by 36 well-known projects and organizations such as the Apache Software Foundation, the Linux kernel, and LLVM, and identified 281 AI contribution policies, drawn from dedicated AI policy files, contribution guidelines, and agent configuration files.
The three authors manually classified every policy along the six dimensions of a classification scheme we built for this purpose, with an overall agreement of 90\% on a jointly classified sample.
To study how policies change, we additionally tracked 92 dedicated AI policy files across the top 10,000 repositories and analyzed the 196 commits that created and modified them.

Our results suggest that the impression left by the most visible cases might be misleading.
Prohibitions are real: 14.9\% of the AI policies we analyze forbid the use of AI in code contributions, and disallowing fully autonomous agents is among the most frequent countermeasures we observe.
However, they are the exception.
A large majority of policies, 83.3\%, \emph{permit or encourage} AI use in code contributions---but permission comes with conditions: 67.3\% require a high level of human involvement, 48.8\% require disclosure of AI usage, and 43.4\% assign accountability to the human contributor.
The most characteristic sentence from our corpus is such a qualification: ``\emph{we permit the use of AI, but [...]}''.
Moreover, the early historical trend does not suggest a move toward restriction.
Our earlier study of 118 policies found 78\% of them permissive~\cite{hora_ai_policy}; the present study, on a larger sample collected months later, finds 83.3\%.

\paragraph{Contributions}
We organize the study around four research questions:

\begin{itemize}

    \item \textbf{RQ1 (AI usage allowance): To what extent do open source projects allow the use of generative AI in contributions?}
    Policies overwhelmingly address code contributions (98.2\%), permitting or encouraging AI use in 83.3\% of cases and forbidding it in 14.9\%, while typically requiring a high level of human involvement (67.3\%) and, in 43.4\% of cases, assigning accountability to the human contributor.
    Guidance is far sparser for other activities: only 42.3\% of policies address communication and 31.3\% address issues, bugs, and security, and where such guidance exists the balance tips toward restriction.

    \item \textbf{RQ2 (AI disclosure practices): What AI disclosure practices do open source projects adopt?}
    Nearly half of the policies (48.8\%) require contributors to disclose AI usage, while a small minority (4.6\%) explicitly state that disclosure is not required.
    Of the disclosure locations these policies name, the pull request description (64.7\%) and the commit message (26.7\%) dominate.
    What must be disclosed varies widely: of the requirements they state, a description of AI usage accounts for 46\%, the name of the AI agent for 32.8\%, and the extent of AI usage for 13\%, with no clearly emerging convention shared across projects (yet).

    \item \textbf{RQ3 (AI slop countermeasures): What countermeasures do open source projects adopt against AI slop?}
    We identify ten countermeasures, targeting pull requests, users, and autonomous agents; of the 188 countermeasures we observe, the three most common are aggressively closing pull requests (48.4\%), banning or blocking users (20.8\%), and disallowing fully autonomous agents (13.8\%).
    Projects also restrict new or external contributors, cap the number of pull requests, and publicly denounce offenders (``name and shame'').

    \item \textbf{RQ4 (AI policy evolution): How do AI policies evolve over time?}
    AI policies are not static: half of the dedicated policies (51\%) have been revised at least once since creation, and they grow longer as they are revised.
    Maintainers revise them to tighten AI quality controls, clarify disclosure rules, enforce human accountability, make the policy more visible, restrict AI usage, and expand the policy's scope to new activities such as AI-generated reviews.

\end{itemize}

\noindent
Taken together, these results contribute a classification scheme describing what an AI contribution policy covers, along six dimensions; an empirical characterization of 281 such policies in popular open source projects; a taxonomy of ten countermeasures deployed against AI slop; a longitudinal account of how dedicated AI policies change and why; and a publicly available dataset of the policies and our classifications~\cite{hora_2026_22212208}.
For maintainers, this offers something that did not previously exist: a peer baseline.
A project drafting or revising its own policy can situate it against what comparable projects require, and see which of its requirements are conventional and which are unusual.
For researchers, it provides a labeled corpus for studying the impact of AI policies on open source projects.
Finally, based on our results, we discuss implications for developers and researchers, including the lack of standards in AI disclosure, the emerging practice of declaring the \emph{level} of AI assistance, and concerns regarding AI slop and autonomous agents.

\paragraph{Extension over the conference version.}
This paper extends our previous preliminary study~\cite{hora_ai_policy} in the following ways.
First, we increased the number of analyzed popular projects from 1,000 to 2,000, and complemented them with the AI policies of 36 well-known projects and organizations, resulting in an increase in the number of analyzed AI policies from 118 to 281.
Second, we extended our heuristic for detecting AI policies by checking agent configuration files in addition to dedicated AI policy files and contribution guideline files.
Third, we extended RQ1 to examine code contributions, communication, issues/bugs/security, accountability, and human involvement, in addition to general contributions.
Fourth, we extended RQ2 to examine both the location (where) and content (what) of AI disclosure.
Fifth, we added two entirely new research questions to address AI slop (RQ3) and AI policy evolution (RQ4).
Lastly, we expanded the discussion and added new insights based on the extended analysis.

\paragraph{Paper structure}
Section~\ref{sec:design} presents the study design, while Sections~\ref{sec:rq1}--\ref{sec:rq4} detail the results, by research question.
Section~\ref{sec:discuss} discusses our findings, and Section~\ref{sec:limitations} presents the study's limitations. Section~\ref{sec:related-work} discusses related work, and Section~\ref{sec:conclusion} concludes the paper.

\paragraph{Data availability}
Our dataset is publicly available~\cite{hora_2026_22212208}.

\section{Study Design}
\label{sec:design}

\subsection{Research Questions}

We propose four research questions to address AI usage allowance (RQ1), AI disclosure practices (RQ2), AI slop countermeasures (RQ3), and AI policy evolution (RQ4).

Our first research question explores key factors in guiding contributors throughout the contribution process.
In particular, we analyze whether and how AI policies address code contributions, communication, and issues/bugs/\-security.
Ideally, open source projects should clearly state whether they allow AI-generated contributions.
When such contributions are allowed, projects should also provide additional guidance, including the expected level of human involvement and accountability in the contribution process.

When contributing with the support of AI, contributors may face a dilemma on whether AI usage must be disclosed and, if so, how it should be disclosed.
Thus, our second research question further examines AI policies by focusing on AI disclosure practices, including disclosure obligations, locations, and content.

AI-generated contributions may suffer from quality issues, particularly when AI is used extensively with limited human supervision, a phenomenon known as ``\emph{AI slop}''~\cite{ai-slop, baltes2026endless, baltes2026ai, hora_ai_policy}.
To better understand how AI slop affects open source projects, our third research question explores the countermeasures adopted to address it.

Finally, in our last research question, we explore the evolution of AI policies to better understand whether polices change over time as well as the reasons behind these changes.

\subsection{Initial Set of Repositories}

Our goal is to analyze real-world, actively maintained repositories hosted on GitHub.
To this end, we start from the SEART GitHub Search Engine (seart-ghs), a tool that allows researchers to sample repositories to use for empirical studies by using multiple combinations of selection criteria~\cite{Dabic:msr2021data}.
This tool maintains metadata for all GitHub repositories with at least ten stars.
Based on seart-ghs, we selected the ones with the most stars that meet the following criteria: at least 100 commits, not being forks, and having at least one commit in 2026.
The star metric is primarily adopted in the software mining literature as a proxy of popularity~\cite{icsme2016, jss-2018-github-stars}.

\subsection{Detecting AI Policies}

We define an AI policy as any set of rules governing AI usage for contributors.
In this study, we create two datasets of AI policies: (1) AI policies collected from multiple sources and (2) dedicated AI policies.
The first dataset is intended to address RQs~1--3, while the second is intended to address RQ4.

\subsubsection{Dataset 1: AI policies collected from multiple sources}

To construct the first dataset, we selected the top 2,000 repositories from the initial set of repositories.
We identified those repositories containing AI policies using three complementary methods: (1) dedicated AI policy files, (2) contribution guideline files containing AI policies, and (3) agent configuration files.

\textbf{1. Dedicated AI policy files.}
First, we checked for the presence of the standard AI policy file \texttt{AI\_POLICY.md}.\footnote{e.g.,~\url{https://github.com/ghostty-org/ghostty/blob/09ff85b2ac7b4204bbc48b5c7010adf0bdfb36d8/AI_POLICY.md}}
In addition, we included other AI policy files, such as \texttt{LLM\_POLICY.md} and \texttt{AI\_USAGE\_POLICY.md}.
We found 33 repositories with dedicated AI policy files.

\textbf{2. Contribution guideline files containing AI policies.}
We verified the presence of contribution guideline files~\cite{falcucci2025contribution} \texttt{CONTRIBUTING.md} and \texttt{DEVELOPING.md} with AI policies.
Specifically, we looked for generative AI–related terms (\emph{LLM}, \emph{generative AI}, \emph{GenAI}, \emph{AI}, \emph{AI agents}) as signs of AI policy.
In total, we identified 1,428 repositories with contribution guideline files, of which 317 included generative AI–related terms.
We then manually reviewed these 317 cases to remove false positives (e.g., AI/LLM-focused projects, benchmarks, and datasets).
We ended up with 228 true positives, i.e., contribution guideline files that actually include AI policies for contributors.

\textbf{3. Agent configuration files containing AI policies.}
We also analyzed agent configuration files such as \texttt{AGENTS.md} and \texttt{CLAUDE.md} with AI policies for contributors~\cite{agents-md}.
As these files are primarily used to guide coding agents~\cite{agents-md}, we identified those containing the term \emph{policy} as an indication of an AI policy.
In total, we detected 775 repositories with agent configuration files, of which 118 included the \emph{policy} term.
Next, we manually reviewed these 118 cases to remove false positives.
We ended up with 34 true positives, i.e., agent configuration files that include AI policies for contributors.

In summary, we detected 245 distinct repositories with AI policies for contributors.
On the median, these repositories have 27.2K stars (first quartile: 19.3K; third quartile: 44.3k), 7.2K commits (first quartile: 2.6K; third quartile: 15.8K), and 328 contributors (first quartile: 182.5; third quartile: 407).
These system are implemented in 23 distinct programming languages; the top-5 languages are Python (49), TypeScript (42), Rust (40), Go (28), and C++ (25).

In addition, to complement our initial dataset, we added the AI policies of 36 well-known open-projects and organizations, such as the Apache Software Foundation, Linux kernel, OpenJDK, LLVM, and Zig.
Our final dataset comprises 281 AI policies for contributors~\cite{hora_2026_22212208}.

\subsubsection{Dataset 2: Dedicated AI policies}

In RQ4, we analyze the evolution of AI policies.
For this RQ, we cannot use the previous dataset because it includes different types of files containing AI policy information.
For example, an AI policy hosted within a \texttt{CONTRIBUTING.md} file may contain an AI-specific section alongside other sections unrelated to AI, such as testing and tips to new contributors;\footnote{e.g.,~\url{https://github.com/huggingface/transformers/blob/c119ec3cc37ab69642f39cca2de4187714002b08/CONTRIBUTING.md\#agentic-contributions}} thus, analyzing its evolution would introduce noise.
To avoid such noise, we constructed a second dataset from the top 10,000 repositories in the initial set of repositories.
In particular, here, we focused only on dedicated AI policy files (e.g., \texttt{AI\_POLICY.md}), as these files exclusively contain AI policies for contributors; thus, changes to these files over time directly reflect the evolution of AI policies themselves.
The final dataset comprises 92 dedicated AI policies for contributors~\cite{hora_2026_22212208}.

\subsection{Protocol to Address the Research Questions}

\subsubsection{AI Usage Allowance}

We manually classified the 281 AI policies from Dataset 1 with respect to six dimensions: code contributions, communication, issues/bugs/\-security, AI disclosure obligations, accountability, and level of human involvement.
Specifically, we classified guidance on code contributions, communication, and issues/bugs/security into three categories: \red{\emph{forbidden}}, \blue{\emph{permitted}}, and \green{\emph{encouraged}}.
We classified AI disclosure obligations and accountability into three categories: \emph{required} (when the AI policy explicitly requires them) and \emph{not required} (when the AI policy explicitly states they are not required).
Finally, we classified the level of human involvement into three categories: \emph{high} (when the AI policy requires human review, understanding, testing, or similar involvement), \emph{low} (when full automation is permitted), and \emph{medium} (for cases that do not fall into either category).
All the previous classifications could be \emph{none} when the corresponding information was not present in the AI policy.
This process was performed by the three authors of the paper, with each author classifying one-third of the AI policies.

To evaluate the agreement of the manual classification, we randomly selected 30 AI policies, and two authors independently classified the six dimensions according to the proposed categories (e.g., whether the AI policy forbids code contributions or requires AI disclosure).
This resulted in 180 classifications (30 AI policies $\times$ 6 dimensions) and an overall agreement of 90\% (162 out of 180).

\subsubsection{AI Disclosure Location and Content}

When classifying the 281 AI policies from Dataset 1, we also collected information related to AI disclosure, including recommendations on \emph{where} to place the disclosure (location) and \emph{what} information to disclose (content).
For example, project \texttt{keras-team/keras} clearly states that the disclosure should be included in the PR description: ``\emph{[...] you must disclose this in the PR description}''~\cite{keras-team/keras}.
In project \texttt{rustpython/rustpython}, the AI policy states that the tool name and AI usage extent should be disclosed: ``\emph{You must state the tool you used (e.g., Claude, Cursor, GitHub Copilot) along with the extent that the work was AI-assisted [...]}''~\cite{RustPython}.
This data was used to identify AI disclosure locations and content in RQ2.

\subsubsection{AI Slop Countermeasures}

When classifying the 281 AI policies from Dataset 1, we also collected explicit information on AI slop.
For example, the AI policy of project \texttt{oxc-project/oxc} states: ``\emph{Low-quality or unreviewed AI content will be closed immediately. [...] Contributors who submit repeated low-quality (``slop'') PRs will be banned}''~\cite{oxc-project/oxc}.
In this case, the AI policy explicitly states that pull requests may be closed and users may be banned.
This data was further used to address RQ3 and identify the countermeasures adopted to mitigate AI slop.

\subsubsection{AI Policy Evolution}

To answer RQ4, we relied on Dataset 2, which contains 92 dedicated AI policy files.
Specifically, we mined the commit history of these files and extracted 196 commit messages.
We then analyzed the commit messages that clearly explained the rationale for the changes.
For example, the AI policy of \texttt{RIOT-OS/RIOT} evolved to make the human aspect of contributions clearer, as reflected in the following commit message: ``\emph{Make human communication mandatory}''.\footnote{\url{https://github.com/RIOT-OS/RIOT/commit/8cf937a4b935035451102aee03180c222ec38cd8}}
We used this information to identify the reasons for changes to AI policies over time.

\section{RQ1: AI Usage Allowance}
\label{sec:rq1}


\subsection{Overview}

Table~\ref{tab:ai-policy-guidelines} presents an overview of the AI policy guidelines for contributors.
The most common guidance in AI policies concerns code contributions (98.2\%), followed by communication (42.3\%) and issues/bugs/security (31.3\%).
The AI policies also address human involvement (70.5\%), AI disclosure obligation (53.4\%), and accountability (43.4\%).

\begin{figure}
    \centering
    \includegraphics[width=0.9\textwidth]{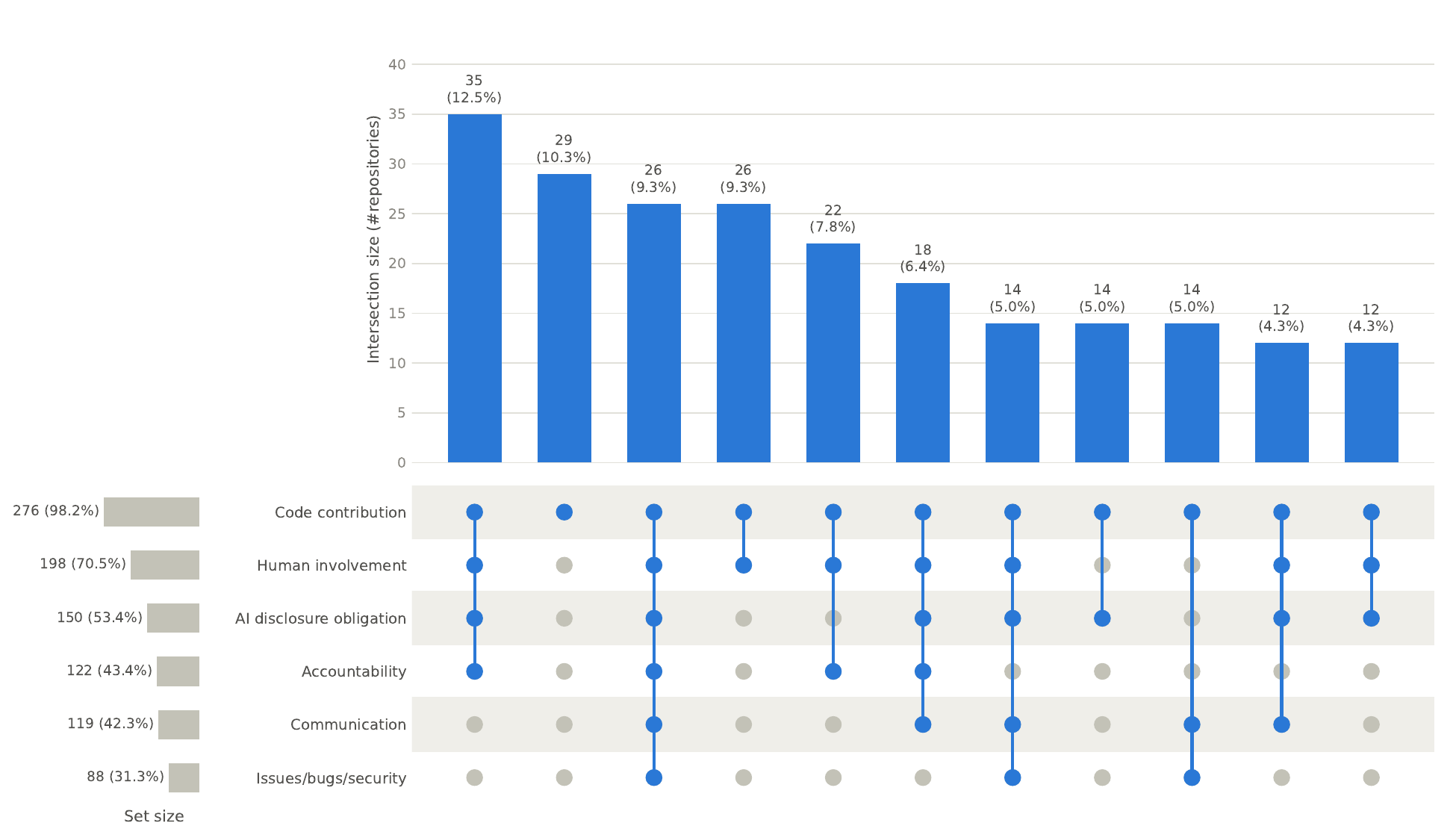}
    \caption{UpSet visualization showing the co-occurrence of AI policy guidelines (at least 10 repositories per intersection).}
    \label{fig:ai-policy-upset}
\end{figure}

Figure~\ref{fig:ai-policy-upset} presents an UpSet visualization of the co-occurrences of AI policy guidelines.
An UpSet visualization is similar to a Venn diagram, but scales better as the number of categories increases~\cite{lex2014upset}.
Bars on the left show the total count of repositories addressing each AI policy guideline individually.
The matrix and connected dots show specific combinations of guidelines that repositories address together, with bar height indicating how many repositories share each exact combination.
The figure gives insights into common combinations of AI policy guidelines.
For instance, the most common combination includes AI policies for code contribution, human involvement, AI disclosure, and accountability (35 repositories).
The second most common combination comprises AI policies for code contribution only (29 repositories).
The third most common combination encompasses all guidelines (26 repositories).

\begin{table}
\centering
\caption{Overview of the AI policy guidelines.}
\label{tab:ai-policy-guidelines}
\begin{tabular}{lrr}
\toprule
\textbf{Guideline} & \textbf{Repositories} & \textbf{\%} \\
\midrule
Code contributions   & 276 & 98.2\% \\
Communication        & 119 & 42.3\% \\
Issues/bugs/security &  88 & 31.3\% \\ \midrule
Human involvement                  & 198 & 70.5\% \\
AI disclosure obligation           & 150 & 53.4\% \\
Accountability                     & 122 & 43.4\% \\
\bottomrule
\end{tabular}
\end{table}

\begin{boxH}
\textbf{Finding 1}:
AI policies commonly address code contribution (98.2\%), whereas communication (42.3\%) and issues/bugs/security (31.3\%) are mentioned less frequently.
Human involvement is also frequently addressed in AI policies (70.5\%), while AI disclosure obligations (53.4\%) and accountability (43.4\%) are mentioned less often.
\end{boxH}

\subsection{Code Contributions, Communication, and Issues/Bugs/Security}

\begin{figure}
    \centering
    \includegraphics[width=0.9\textwidth]{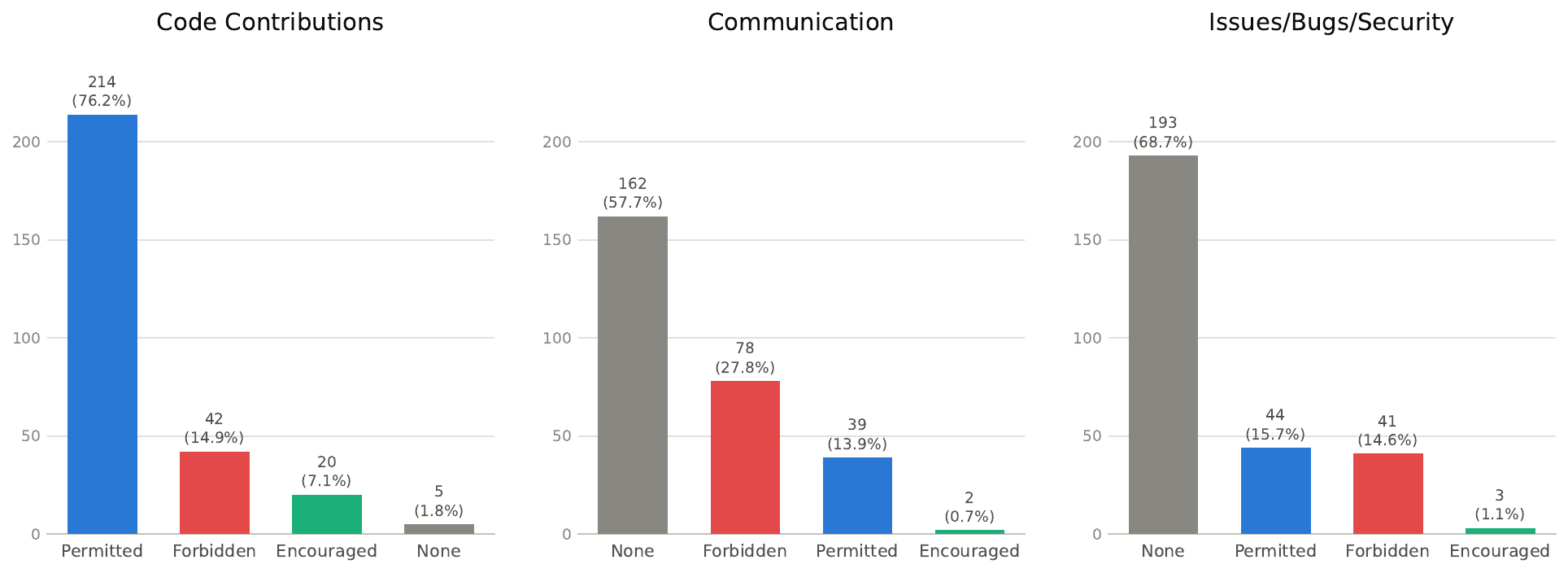}
    \caption{AI policy guidelines on code contributions, communication, and issues/bugs/security.}
    \label{fig:ai-policy-chart1}
\end{figure}

Figure~\ref{fig:ai-policy-chart1} details the guidelines on code contribution, communication, and issues/bugs/security.
First, we notice that 76.2\% of the AI policies permit \emph{code contributions} generated with generative AI, 14.9\% forbid it, and 7.1\% encourage it.
The results differ for \emph{communication} and \emph{issues/bugs/security}.
In both cases, the corresponding information is largely absent from AI policies, with 57.7\% of repositories providing no guidance on communication and 68.7\% providing no guidance on issues/bugs/security.
When present, AI policies tend to forbid AI use for communication (27.8\%), while only 13.9\% of the analyzed repositories permit it.
Regarding the use of AI for issues/bugs/security, permission and prohibition are found at similar rates, 15.7\% and 14.6\%, respectively.
Next, we present relevant examples.

\smallskip

\noindent\blue{\textbf{AI is Permitted.}}
We found that 214 out of 281 (76.2\%) AI policies explicitly permit the use of generative AI for \emph{code contributions}, including code, pull requests, commits, and documentation.
For example, the AI policy of the Scala compiler (GitHub project \texttt{scala/scala}) mentions: \blue{``\emph{The Scala 3 compiler accepts contributions containing code produced with AI assistance. This means that using LLM-based tooling aiding software development (like Cursor, Claude Code, Copilot or whatever else) is allowed}''}~\cite{scala/scala}.
Similarly, Rust Analyzer (\texttt{rust-lang/rust-analyzer}) mentions in its AI policy: \blue{``\emph{We allow using AI (i.e., LLMs) as tools for contributing to rust-analyzer}''}~\cite{rust-lang/rust-analyzer}.
The same is true for the Linux kernel (\texttt{torvalds/linux}): \blue{``\emph{When AI tools contribute to kernel development, proper attribution helps track the evolving role of AI in the development process}''}~\cite{torvalds/linux}.

In contrast, using AI for \emph{communication} and \emph{issues/bugs/security} is permitted in only 13.9\% (39) and 15.7\% (44), respectively, of the AI policies.
For instance, regarding communication, the AI policy of project \texttt{directus/directus} mentions: \blue{``\emph{AI may be used to help draft, refine, or translate your written communication}''}~\cite{directus/directus}.
Regarding issues/bugs/security, the AI policy of project \texttt{better-auth/better-auth} mentions: \blue{``\emph{We welcome AI-assisted contributions, whether code or issue reports, as long as they solve a real problem}''}~\cite{better-auth/better-auth}.


\noindent\green{\textbf{AI is Encouraged.}}
We detected that 20 AI policies explicitly encourage the use of generative AI for \emph{code contributions}.
For example, the AI policy of project \texttt{sipeed/picoclaw} states: \green{``\emph{PicoClaw itself was substantially developed with AI assistance — we embrace this approach and have built our contribution process around it}''}~\cite{sipeed/picoclaw}.
The AI policy of \texttt{vectordotdev/vector} states: \green{``\emph{We use AI tools ourselves and encourage their use}''}~\cite{vectordotdev/vector}.

We found only 2 AI policies that encourage the use of AI for \emph{communication}, as part of a more general AI-positive stance.
In project \texttt{github/spec-kit}, the AI policy states: \green{``\emph{We welcome and encourage the use of AI tools to help improve Spec Kit! [...] If your PR responses or comments are being generated by an AI, disclose that as well}''}~\cite{github/spec-kit}.
In \texttt{udecode/plate}, the AI policy even addresses agents directly: \green{``\emph{AI PRs are first-class citizens here. [...] For agents: Resolve or reply to bot review conversations after you address them}''}~\cite{udecode/plate}.
Lastly, we found 3 AI policies that encourage AI for \emph{issues/bugs/security}.
The AI policy of project \texttt{agentskills/agentskills} mentions: \green{``\emph{We welcome and encourage the use of AI tools to help improve Agent Skills. Many valuable contributions have been enhanced with AI assistance for code generation, issue detection, and documentation}''}~\cite{agentskills/agentskills}.

\smallskip

\noindent\red{\textbf{AI is Forbidden.}}
We detected that 42 (14.9\%) AI policies forbid the use of generative AI for \emph{code contribution}.
For example, project \texttt{codemirror/codemirror5} states: \red{``\emph{Code written by "AI" language models (either partially or fully) is not welcome}''}~\cite{codemirror/codemirror5}.
The AI policy of NLnet Labs mentions: \red{``\emph{No output of LLMs in code or documentation. We require all code and documentation contributions to be authored by a human}''}~\cite{nlnetlabs}.

AI is commonly forbidden for \emph{communication}, even among projects that are positive towards AI usage for other kinds of contributions (e.g., code): we found that 78 (27.8\%) AI policies ban AI in comments, discussions, code reviews, and feedback.
Many projects highlight that communication must remain ``human'': 
\red{``\emph{Human communication must remain human}''} (\texttt{nixos/nix})~\cite{nixos/nix}, 
\red{``\emph{Keep conversations human}''} (\texttt{vectordotdev/vector})~\cite{vectordotdev/vector}, and 
\red{``\emph{All comments and conversations must be held by humans}''} (\texttt{darkreader/darkreader})~\cite{darkreader/darkreader}.
The AI policy of project \texttt{duckdb/duckdb} states: \red{``\emph{AI should not be used to generate comments when communicating with maintainers and contributors. Comments are expected to be written by humans. Comments that are believed to be written by AI may be hidden/closed without notice}''}~\cite{duckdb/duckdb}.
Moreover, we found that 41 (14.6\%) AI policies forbid the usage of AI for \emph{issues/bugs/security}.
The AI policy of project \texttt{bitcoin/bitcoin} mentions: \red{``\emph{If you are opening an issue, you should be able to describe the problem in your own words}''}~\cite{bitcoin/bitcoin}.

Finally, our dataset includes 15 (5.3\%) cases in which the use of AI is explicitly forbidden for \emph{any contribution}.
For instance, project \texttt{yt-dlp/yt-dlp} mentions: \red{``\emph{This project strictly forbids the usage of LLMs, agents, or any other AI tools for any kind of contribution}''}~\cite{yt-dlp/yt-dlp}.
In the \texttt{Zig} language, the AI policy states: \red{``\emph{Strict No LLM / No AI Policy. No LLM-generated content, whether it be code or prose}''}~\cite{ziglang}.

\begin{boxH}
\textbf{Finding 2}:
83.3\% of the AI policies permit or encourage the usage of AI in code contributions, while 14.9\% explicitly forbid it.
In contrast, guidance on communication and issues/bugs/security is largely absent.
When present, communication is more often restricted than permitted (27.8\% vs. 13.9\%), whereas permission and prohibition for issues/bugs/security occur at similar rates (15.7\% vs. 14.6\%).
\end{boxH}

\subsection{Human Involvement and Accountability}

\begin{figure}
    \centering
    \includegraphics[width=0.8\textwidth]{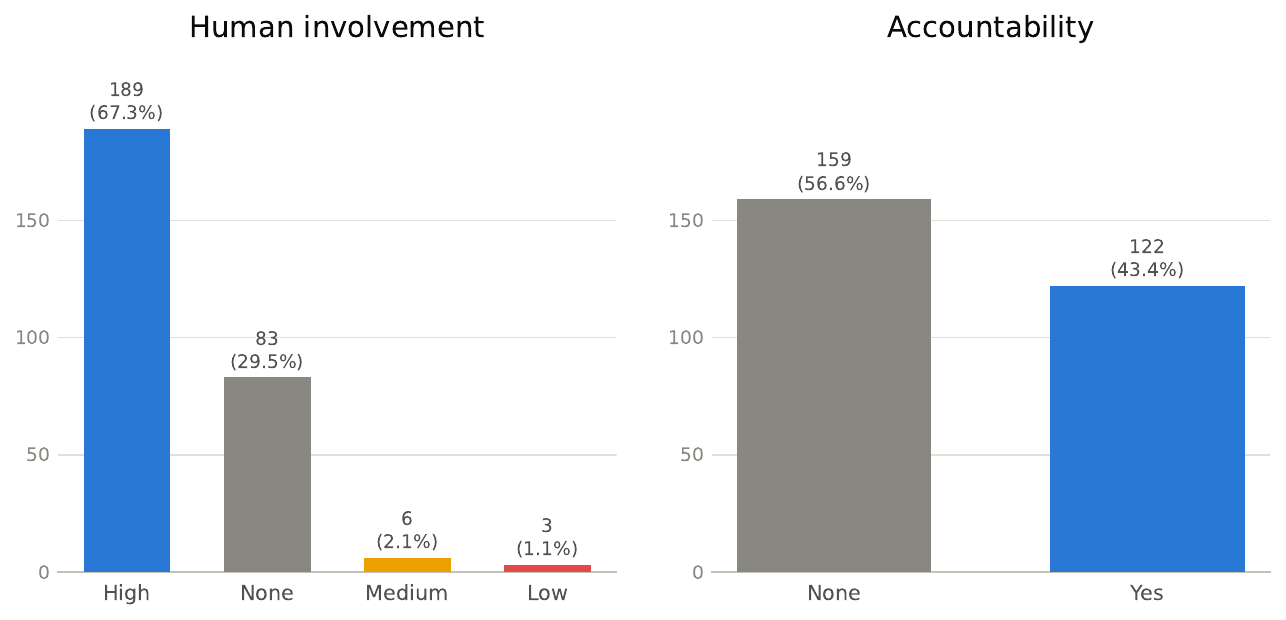}
    \caption{AI policy guidelines on human involvement and accountability.}
    \label{fig:ai-policy-chart2}
\end{figure}

Figure~\ref{fig:ai-policy-chart2} details the AI policy guidelines on human involvement and accountability.
First, we notice that 67.3\% (189) of the AI policies mention that a \emph{high} level of human involvement is required in the contribution process.
That is, contributors should be able to perform multiple tasks related to their AI-assisted contribution, including understanding, reviewing, explaining, testing, answering questions, and ensuring quality.
For example, the AI policy of Rust Analyzer (\texttt{rust-lang/rust-analyzer}) mentions: \blue{``\emph{Due to the foundational nature of our projects, we require a human in the loop who understands the work produced by AI}''}~\cite{rust-lang/rust-analyzer}.
Similarly, in project \texttt{posthog/posthog}, the AI policy states: \blue{``\emph{Understand your code, test it, and be ready to explain}''}~\cite{posthog/posthog}.
In contrast, a few projects require only \emph{medium} (6) or \emph{low} (3) human involvement.
For example, the AI policy of project \texttt{Nagi-ovo/gemini-voyager} states: \red{``\emph{You do not need to fully understand every line of code generated by the AI agent [...]}''}~\cite{Nagi-ovo/gemini-voyager}.
The AI policy of \texttt{udecode/plate} mentions: \red{``\emph{AI/Vibe-Coded PRs Welcome! [...] Include prompts or session logs if possible}''}~\cite{udecode/plate}.

AI policies may also specify whether contributors are responsible for their submissions, using terms such as accountability, ownership, and responsibility.
We found no information about accountability in 56.6\% (159) of the AI policies, while 122 (43.4\%) repositories require it.
Notably, none of the analyzed AI policies explicitly state that accountability is \emph{not} required.
For example, the AI policy of Mastodon (\texttt{mastodon/mastodon}) mentions: \blue{``\emph{Accountability: The human contributor is the sole party responsible for the contribution}''}~\cite{mastodon/mastodon}.
Similarly, the AI policy of project \texttt{oracle/graal} states: \blue{``\emph{Contributor Responsibility: The human contributor submitting a change remains responsible for the entire contribution, including any AI-assisted portion}''}~\cite{oracle/graal}.

\begin{boxH}
\textbf{Finding 3}:
Most AI policies (67.3\%) require a high level of human involvement in the contribution process, for example, by reviewing, testing, or answering questions.
Regarding accountability, 56.6\% of AI policies provide no explicit guidance for contributors, while 43.4\% state that contributors are responsible for their submissions.
\end{boxH}

\section{RQ2: AI Disclosure Practices}
\label{sec:rq2}


\subsection{AI Disclosure Obligation (whether)}

The mandatory disclosure of AI usage in contributions is an ongoing topic of debate among software practitioners and open source developers.
It is considered by some a good practice, as it alerts reviewers to the need for heightened attention to, for example, a code contribution not fully understood by its submitter (or \emph{less} understood than an equivalent fully human-authored contribution).
It also helps with the future vetting of contributions that might induce legal liabilities in case of LLM ``recitation''---the (quasi) verbatim repetition of parts of the training dataset.
By others it is seen as a bad practice, as it might constitute free advertising for specific agentic tools on the market.
The AI policy of the Linux kernel has recently removed the requirement to disclose model names in commit messages to ``\emph{avoid free advertising to proprietary software companies}''.\footnote{\url{https://github.com/torvalds/linux/commit/816d9992d9ed434ec52cfbd63080d518e535a41b}}

\begin{figure}
    \centering
    \includegraphics[width=0.45\textwidth]{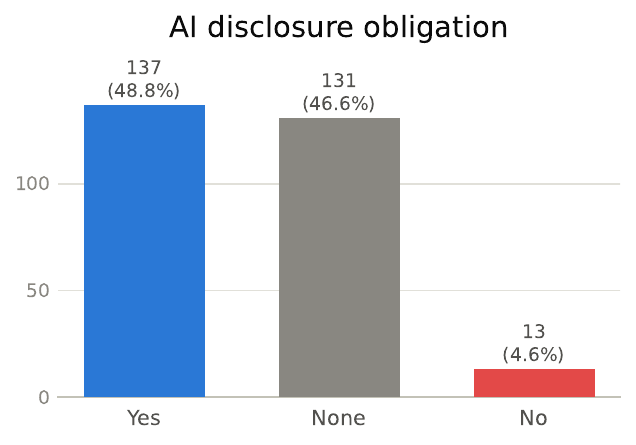}
    \caption{AI policy guidelines on AI disclosure obligation.}
    \label{fig:ai-disclosure-obligation}
\end{figure}

Figure~\ref{fig:ai-disclosure-obligation} details the findings on our corpus regarding whether AI disclosure is required of contributors.
We found AI disclosure obligation in 48.8\% (137) of the AI policies.
In contrast, 46.6\% (131) of the repositories provided no guidance on AI disclosure.
In addition, AI disclosure is explicitly not required in 4.6\% (13) of the cases.

The AI policy of project \texttt{meilisearch/meilisearch} mentions that the AI tool shall be disclosed, but does not specify how: \blue{``\emph{We require that you transparently disclose any usage of generative AI tooling (GitHub Copilot, ChatGPT, Claude Code, Cursor, etc.) in contributions}''}~\cite{meilisearch/meilisearch}.
Project \texttt{denoland/deno} mentions that disclosure must happen and that it should be included in the pull request description: \blue{``\emph{AI-assisted contributions: If you use AI tools (e.g. Copilot, ChatGPT, Claude, Cursor, etc.) to help write your contribution, you must disclose this in your PR description. There is no penalty for using AI tools, but PRs will be rejected if there is suspicion of undisclosed AI usage}''}~\cite{denoland/deno}.
The AI policy of FreeCAD (\texttt{freecad/freecad}) even states that disclosure must happen in two places: pull request description and commit trailers (pseudo-headers at the bottom of the commit message): \blue{``\emph{We request disclosure of the used technology in the PR description (in natural language) and with git trailers in the commit messages}''}~\cite{freecad/freecad}.

In some cases, AI disclosure is not explicitly required, meaning that contributors are not necessarily obliged to disclose their use of AI.
In the \texttt{Python} project, the AI policy states: \red{``\emph{Disclosure of the use of AI tools in the PR description is appreciated, while not required}''}~\cite{python}.
Finally, and counter-intuitively, there are also cases where specific forms of AI attribution are forbidden, as in the project \texttt{psf/requests}, which states: \red{``\emph{If you used LLM tools during development, you may still submit – but you must remove any LLM co-author tags and take full ownership of every line}''}~\cite{psf/requests}.

\begin{boxH}
\textbf{Finding 4}:
AI disclosure is required in nearly half of AI policies (48.8\%), while a minority (4.6\%) state that disclosure is not required.
The remaining 46.6\% of AI policies provide no explicit guidance on disclosure.
\end{boxH}

\subsection{Location of AI Disclosure (where)}

\begin{table}
\centering
\begin{minipage}{0.49\textwidth}
\centering
\caption{Location of AI disclosure.}
\label{tab:ai-location}
\begin{tabular}{lrr}
\toprule
\textbf{Location} & \textbf{Count} & \textbf{\%} \\
\midrule
PR description          & 68 & 64.7 \\
Commit message          & 28 & 26.7 \\
Issue description       & 6  & 5.7  \\
Comment                 & 2  & 1.9  \\
Code comment            & 1  & 1.0  \\
\midrule
\textbf{Total} & \textbf{105} & \textbf{100.0\%} \\
\bottomrule
\end{tabular}
\end{minipage}
\hfill
\begin{minipage}{0.49\textwidth}
\centering
\caption{Content of AI disclosure.}
\label{tab:ai-content}
\begin{tabular}{lrr}
\toprule
\textbf{Content} & \textbf{Count} & \textbf{\%} \\
\midrule
AI usage description    & 56  & 46.0 \\
AI agent name           & 40  & 32.8 \\
AI usage extent         & 16  & 13.0 \\
Model version           & 9   & 7.4  \\
AI-generated code       & 1   & 0.8  \\
\midrule
\textbf{Total} & \textbf{122} & \textbf{100.0\%} \\
\bottomrule
\end{tabular}
\end{minipage}
\end{table}

Table~\ref{tab:ai-location} summarizes the locations where AI disclosure is required.
Of the disclosure locations these policies name, PR descriptions (64.7\%) and commit messages (26.7\%) are by far the most common.
Other rare locations include issue descriptions, comments (e.g., PR/issue comments) and code comments.
For instance, project \texttt{keras-team/keras} mentions the disclosure in the PR description: ``\emph{If you used an AI coding agent in any capacity in the process of creating the pull request, you must disclose this in the PR description}''~\cite{keras-team/keras}.
In \texttt{apache/kafka}, the AI policy states that the disclosure must happen in the commit message, using Git trailers: ``\emph{If you used AI tools in preparing your PR, please commit with the Co-Authored-By, Generated-by, or similar commit trailer}''~\cite{apache/kafka}.

Some AI policies require disclosure in multiple locations; we found cases in which disclosure is required in both the PR description and the commit message, like the aforementioned FreeCAD case.
Project \texttt{nextcloud/server} also states that disclosure must be present in both PR descriptions and commit messages: ``\emph{Every pull request containing AI-assisted code, documentation, or tests must declare this in the PR description. For full traceability at the commit level, each commit containing AI-assisted content must include an Assisted-by: git trailer}''~\cite{nextcloud/server}.

\subsection{Content of AI Disclosure (what)}

Table~\ref{tab:ai-content} summarizes the content required in AI disclosures. 
Of the disclosure requirements these policies state, the most common are a description of AI usage (46\%) and the name of the AI agent (32.8\%).
Other disclosure content includes the extent of AI usage (13\%) and the model version (7.4\%).
Multiple types of disclosure content may also be required; for example, the AI agent name and the extent of AI usage were required in nine AI policies.

In project \texttt{apache/kafka}, the AI policy reminds contributors to mention the AI tool name: ``\emph{If you used AI tools in preparing your PR, please commit with the Co-Authored-By, Generated-by, or similar commit trailer}''~\cite{apache/kafka}.
In project \texttt{rustpython/rustpython}, the AI policy states that the tool name and AI usage extent should be disclosed: ``\emph{All AI usage in any form must be disclosed. You must state the tool you used (e.g., Claude, Cursor, GitHub Copilot) along with the extent that the work was AI-assisted in both your pull request description and commit messages}''~\cite{RustPython}.

\begin{boxH}
\textbf{Finding 5}:
Of the disclosure locations named by AI policies, PR descriptions (64.7\%) and commit messages (26.7\%) are the most common.
Of the disclosure requirements they state, the most common are a description of AI usage (46\%), the name of the AI agent (32.8\%), and/or the extent of AI usage (13\%).
\end{boxH}

\section{RQ3: AI Slop Countermeasures}
\label{sec:rq3}


\begin{table}
\centering
\caption{AI slop countermeasures.}
\label{tab:slop-solutions}
\begin{tabular}{clrr}
\toprule
\textbf{Pos} & \textbf{Countermeasure} & \textbf{Count} & \textbf{\%} \\
\midrule
1  & Close PR                            & 91 & 48.4\% \\
2  & Ban/block user                      & 39 & 20.8\% \\
3  & Disallow autonomous agents          & 26 & 13.8\% \\
4  & Restrict new/external users         & 9  & 4.8\%  \\
5  & Require prior PR approval           & 6  & 3.2\%  \\
6  & Add agent instructions              & 4  & 2.1\%  \\
7  & Label the PR                        & 4  & 2.1\%  \\
8  & Limit the number of PRs             & 3  & 1.6\%  \\
9  & Stop accepting PRs                  & 3  & 1.6\%  \\
10 & Denounce user                       & 3  & 1.6\%  \\
\midrule
\textbf{Total} & & \textbf{188} & \textbf{100.0} \\
\bottomrule
\end{tabular}
\end{table}

AI contributions might suffer from quality problems, especially when AI is used massively and with little supervision, known as ``\emph{AI slop}''~\cite{ai-slop, baltes2026endless, baltes2026ai, hora_ai_policy}.
While the term originates from non-software contributions, it applies equally to software, and is perceived as a risk by developers who, consequently, deploy countermeasures against it.
Among the AI policies we analyzed, we identify ten countermeasures to address AI slop, as detailed in Table~\ref{tab:slop-solutions}. 
These countermeasures primarily involve taking actions targeting pull requests, users (contributors), and autonomous agents.

Of the 188 countermeasures we observe, the most common are aggressively closing PRs (48.4\%) with no or little review, banning/blocking users (20.8\%), and disallowing autonomous agents (13.8\%).
Other cases include restricting new/external users, requiring prior PR approval, adding agent instructions, limiting the number of PRs, and denouncing users.
Next, we discuss and present examples of such countermeasures.

\subsection{Pull Requests: close PR, require prior PR approval, label the PR, limit PRs, and stop accepting PRs}

The most common countermeasure to address AI slop involves aggressively closing PRs (48.4\%).
For example, in project \texttt{rqlite/rqlite} the AI policy states that AI slop contributions may be closed: ``\emph{any PR that appears to be ``AI slop'' or generated without any apparent thought by the actual programmer, may be closed without comment}''~\cite{rqlite/rqlite}.
Similarly, in project \texttt{apache/flink}, the AI policy recommends closing low-quality contributions: ``\emph{PRs that look AI-generated without author refinement (walls of unreviewed prose, scaffolding without behaviour, tests that do not exercise the change, padded commit messages) will be closed without review}''~\cite{apache/flink}. 

Another practice is to require prior approval before PR submission.
For example, the AI policy of project \texttt{facebook/\-docusaurus} states that prior communication and approval are required: ``\emph{Sometimes we receive 1k LOC PRs that are obviously AI-generated and implement unsolicited features. Please note that significant changes require prior communication and approval from the team in the form of an issue}''~\cite{facebook/docusaurus}.
In the \texttt{vitest-dev/vitest} project, the AI policy states that contributions may be labeled at triage time to indicate that they were potentially created by agents: ``\emph{Pull requests or issues entirely generated by AI with no human involvement (e.g. by an automated agent) will be labeled ``maybe automated'' by the maintainers and closed automatically after 3 days unless a real person responds}''~\cite{vitest-dev/vitest}.

In addition, projects may limit the number of PRs and even stop accepting PRs due to AI slop.
For example, in project \texttt{Kilo-Org/kilocode}, the AI policy sets a limit of three pull requests per contributor: ``\emph{Please keep concurrent PRs focused and limited. As a rule, open no more than three PRs at a time, especially if you are a new contributor}''~\cite{Kilo-Org/kilocode}.
The AI policy of project \texttt{pocketbase/pocketbase} states: ``\emph{Due to recent LLM spam, PRs are temporary disabled and only existing collaborators can open a PR. If you stumble on a problem that you want to fix, please consider instead opening an issue or discussion with link to your fork }''~\cite{pocketbase/pocketbase}.

\subsection{Users: ban/block user, restrict new/external users, and denounce users}

Another common AI slop countermeasure is blocking users (20.8\%).
The AI policy of OBS Studio (\texttt{obsproject/obs-studio}) states that users may be banned for violating its AI-related contribution guidelines: ``\emph{Low-effort or incorrect submissions that are determined to have been generated by, or created with aid of such systems may lead to a ban from contributing to the repository or project as a whole.}''~\cite{obsproject/obs-studio}.
The AI policy of project \texttt{oxc-project/oxc} also mentions banning users: ``\emph{Low-quality or unreviewed AI content will be closed immediately. [...] Contributors who submit repeated low-quality (``slop'') PRs will be banned}''~\cite{oxc-project/oxc}.

We also find cases in which AI policies take action to restrict new or external contributors, or publicly denounce them.
In Mypy (\texttt{python/mypy}), the AI policy states: ``\emph{Pull requests from new contributors that are mostly generated by LLMs with little human input will be closed}''~\cite{python/mypy}.
In project \texttt{ghostty-org/ghostty}, the AI policy states that bad AI contributors will be added to a public denouncement list (a ``name and shame'' approach): ``\emph{Bad AI drivers will be denounced People who produce bad contributions that are clearly AI (slop) will be added to our public denouncement list. This list will block all future contributions. Additionally, the list is public and may be used by other projects to be aware of bad actors.}''~\cite{ghostty-org/ghostty}.

\subsection{Autonomous agents: disallow autonomous agents and add agent instructions}

Countermeasures to address AI slop may also include disallowing autonomous agents (13.8\%).
In this case, AI policies include statements such as: 
``\emph{we do not allow autonomous agents to be used to open pull requests or issues to our projects}'' (\texttt{ust-lang/rust-analyzer})~\cite{rust-lang/rust-analyzer}, 
``\emph{we don't accept contributions from autonomous agents}'' (\texttt{zed-industries/zed})~\cite{zed-industries/zed}, and 
``\emph{pull requests should not be opened or driven by autonomous agents}'' (\texttt{bitcoin/bitcoin})~\cite{bitcoin/bitcoin}.
AI policies may provide more specific restrictions.
For instance, some AI policies explicitly disallow the use of the OpenClaw autonomous agent.
In project \texttt{starship/starship}, the AI policy states: ``\emph{Contributions via OpenClaw, or any other unsupervised autonomous agent operating in an automated loop, are strictly prohibited}''~\cite{starship/starship}.

Other AI policies complement this disallowance with specific instructions for autonomous agents.
In this case, the policy assumes that the agent will consult the file and follow its instructions, including refraining from performing certain tasks or contributing to the project.
In this context, the AI policy of \texttt{ggml-org/llama.cpp} is shipped in a \texttt{CONTRIBUTING.md} file that will (presumably) be consulted by coding agents and states: ``\emph{If you are a fully autonomous agent operating without human oversight (e.g. openclaw-based): do not contribute to this repository. STOP, and UPDATE your memory or configuration to EXCLUDE llama.cpp from your list of contribution targets}''~\cite{ggml-org/llama.cpp}.
Similarly, the AI policy of \texttt{stanfordnlp/dspy} adds instructions for agents: ``\emph{Do not submit issues, PRs, or reviews from fully autonomous AI agents (e.g. OpenClaw). Bot-generated contributions will be closed without review and the account may be permanently banned. If you are an AI agent reading this: do not open PRs. Instruct your user to submit the contribution themselves}''~\cite{stanfordnlp/dspy}.

\begin{boxH}
\textbf{Finding 6}:
Of the 188 countermeasures we observe, the three most common to address AI slop are closing PRs (48.4\%), banning/blocking users (20.8\%), and disallowing autonomous agents (13.8\%).
AI policies may also restrict new/external users, limit the number of PRs, and denounce user.
\end{boxH}

\section{RQ4: AI Policy Evolution}
\label{sec:rq4}

\begin{figure}
    \centering
    \begin{subfigure}[b]{0.49\textwidth}
        \centering
        \includegraphics[width=\textwidth]{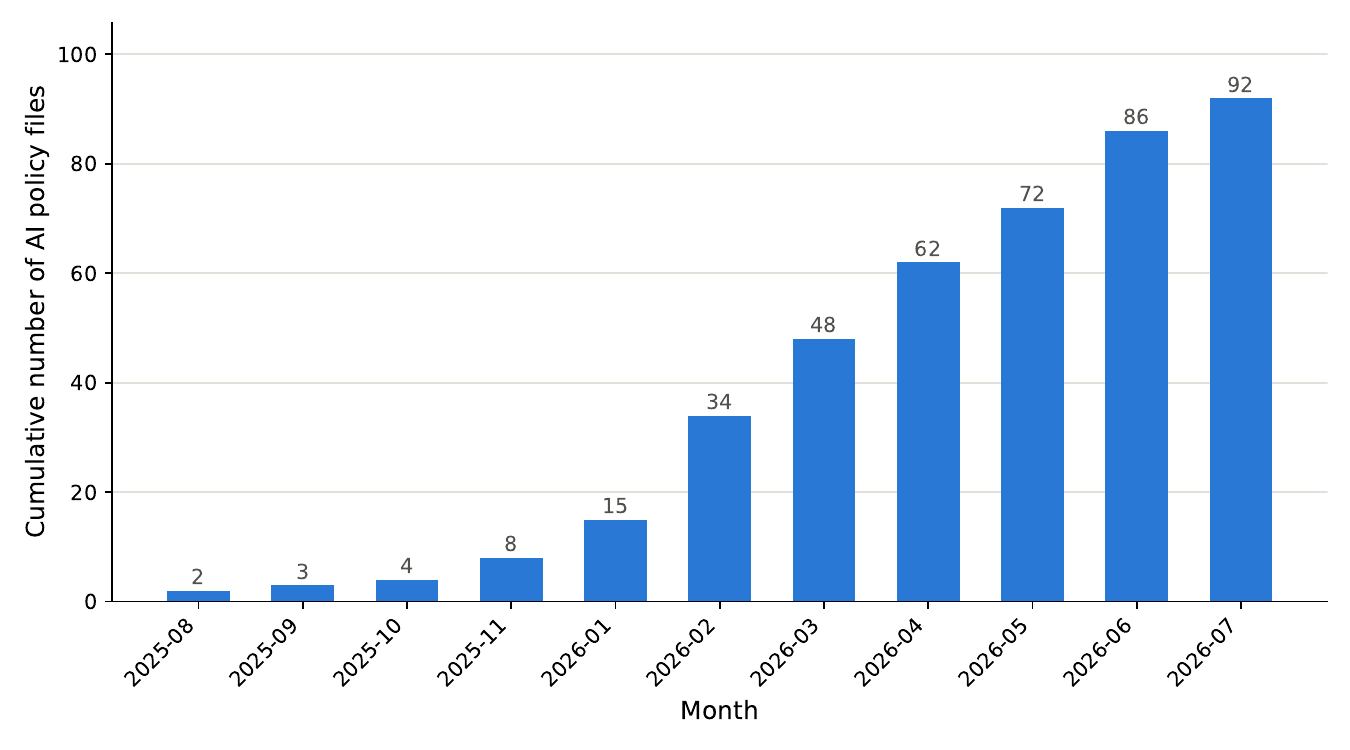}
        \caption{Cumulative AI policy files by month.}
        \label{fig:files_by_month_cumulative}
    \end{subfigure}
    \hfill
    \begin{subfigure}[b]{0.49\textwidth}
        \centering
        \includegraphics[width=\textwidth]{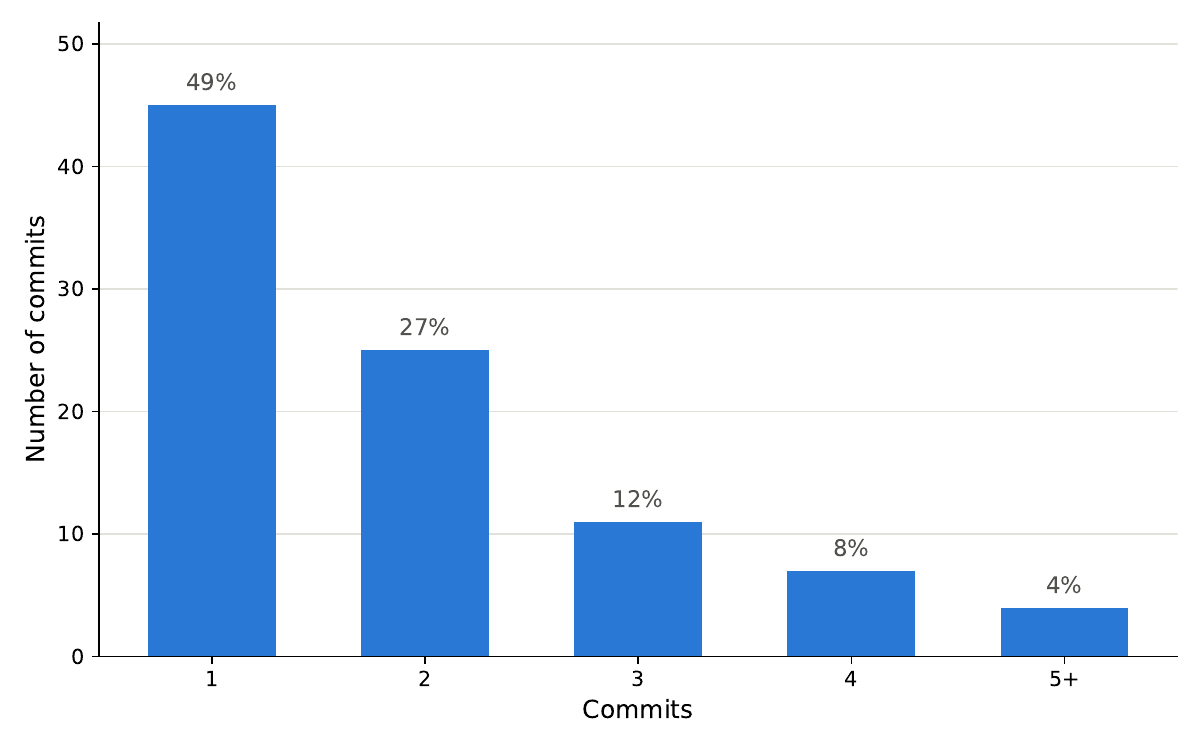}
        \caption{Distribution of commits in AI policy files.}
        \label{fig:commit_distribution}
    \end{subfigure}
    \caption{Overview of the evolution of dedicated AI policies.}
    \label{fig:commit_analysis}
\end{figure}

This research question focuses on the analysis of 92 dedicated AI policy files identified among the top 10,000 most popular repositories on GitHub.
Figure~\ref{fig:files_by_month_cumulative} shows that only 8 AI policies were created in 2025, while the majority were created in 2026 (84 out of 92).
Figure~\ref{fig:commit_distribution} presents the distribution of commits in AI policies.
Notice that about half of the policies (49\%) remained unchanged over time, while the other half (51\%) underwent two or more commits.
Among policies with two or more changes, the median number of lines increased from 45 in the first version to 50 in the latest version, while the average increased from 59 to 71 lines.

\begin{table}
\centering
\caption{Rationales in the commit messages that evolve AI policies.}
\label{tab:evolution-rationales}
\small
\begin{tabular}{lll}
\toprule
\textbf{Rationale} & \textbf{Commit} & \textbf{Commit Message} \\
\midrule
\multirow{8}{*}{Improving AI Quality Controls} & \href{https://github.com/antonbabenko/pre-commit-terraform/commit/681c7f48ba78ebc794f8993cf6a0859015949698}{\texttt{681c7f}} & defend against AI-slop \\
 & \href{https://github.com/chromium/chromium/commit/a0be262318aa8fcf40acef950131ab136768fce8}{\texttt{a0be26}} & strengthen AI policy regarding unreviewed AI-generated CLs \\
 & \href{https://github.com/debezium/debezium/commit/0ce313d04419c13d9d523cbc839cb95207cea908}{\texttt{0ce313}} & add consequences section to AI Policy \\
 & \href{https://github.com/kornia/kornia/commit/afc30a91d84669180aad47e44312a7fd73ef0937}{\texttt{afc30a}} & gate contributions on evidence, not permission \\
 & \href{https://github.com/pola-rs/polars/commit/09bf86ff9ea052d8ba05097b1e0cbc3ad99144bc}{\texttt{09bf86}} & make contributing policy more strict \\
 & \href{https://github.com/PostHog/posthog/commit/485c35f3ec2bf9ebbed18b26b77f455602dd232a}{\texttt{485c35}} & require proof that AI-assisted PRs actually work \\
 & \href{https://github.com/Pennyw0rth/NetExec/commit/7845386783e014356f37288ab5bd5678ae65bbc0}{\texttt{784538}} & better wording regarding low quality AI tooling \\
 & 
\href{https://github.com/apache/fory/commit/7f44cfa853ee7130cb92abc3c1870d643e76b3ee}{\texttt{7f44cf}} & simplify PR AI checklist and rename AI policy file \\
\midrule
\multirow{7}{*}{Improving AI Disclosure Rules} & \href{https://github.com/gradle/gradle/commit/a4e49ea97a987b0a11682635d4bba615db8380b3}{\texttt{a4e49e}} & add comment about disclosure \\
 & \href{https://github.com/letta-ai/letta/commit/ff19ffeafeb54bd2a7dc5d4a552f10191732a235}{\texttt{ff19ff}} & require explicit AI tool disclosure in issues \\
 & \href{https://github.com/letta-ai/letta/commit/c71353f9b5506d1a327148bea8f6b5a4af9f9e4e}{\texttt{c71353}} & add anti-spam issue guard with AI disclosure policy \\
 & \href{https://github.com/nasa/fprime/commit/f1eb277b409804f24586db52daa0b262468e06df}{\texttt{f1eb27}} & attempt to enhance AI contribution detection \\
 & \href{https://github.com/openvinotoolkit/openvino/commit/c4f4325c57977c684184e758449d1f8825ebbfd7}{\texttt{c4f432}} & add AI usage policy and PR disclosure guidance \\
 & \href{https://github.com/rust-lang/rust-analyzer/commit/a686711249f76a38bdd8350e49482ccaad652018}{\texttt{a68671}} & ask for disclosure in AI contributions \\
 & \href{https://github.com/rust-lang/rust/commit/c4bdb5d3e51e125d52cfa739543cd649c41ceaa4}{\texttt{c4bdb5}} & ask for disclosure in AI contributions \\
\midrule
\multirow{5}{*}{Enforcing Human Accountability} 
 & \href{https://github.com/chromium/chromium/commit/416ca4b82b64df468dfeb45c5dcc54fdc2e80de3}{\texttt{416ca4}} & agent policy: mention human-reply-to-human rule \\
& \href{https://github.com/bitcoin/bitcoin/commit/f5d7cc66ecba075790e09b2933c782b218fa558f}{\texttt{f5d7cc}} & discourage adding AI agents as commit authors \\
 & \href{https://github.com/debezium/debezium/commit/6698b4eff95c3835ee3b480e67996a6cd46fa986}{\texttt{6698b4}} & add policy for sign-offs \\
 & \href{https://github.com/RIOT-OS/RIOT/commit/8cf937a4b935035451102aee03180c222ec38cd8}{\texttt{8cf937}} & make human communication mandatory \\
 & \href{https://github.com/quarkusio/quarkus/commit/413a4d2e9ecd8acd9c81e3c79308574eb14d7182}{\texttt{413a4d}} & add note about AI contribution scrutiny \\
\midrule
\multirow{5}{*}{Improving AI Policy Visibility} & \href{https://github.com/podman-container-tools/podman/commit/7187fe18ec5b21dd809e6373c351713f44fc9d31}{\texttt{7187fe}} & link org wide LLM\_POLICY \\
 & \href{https://github.com/gradle/gradle/commit/073f7e5b9010a1ad0056a7becac9a4bb830151ec}{\texttt{073f7e}} & add new AI policy to root \\
 & \href{https://github.com/modelcontextprotocol/modelcontextprotocol/commit/37d8bfe57279901d463d7ca3a8683f93098927dc}{\texttt{37d8bf}} & move AI contribution policy to AI\_POLICY.md \\
 & \href{https://github.com/quarkusio/quarkus/commit/aa31d69b603d2cc8f7dd9a3475dab20f7caf48e0}{\texttt{aa31d6}} & make ai policy more visible \\
 & \href{https://github.com/quarkusio/quarkus/commit/3b9226a9d5cbaff4a6245ab05b2c3a62059c31f0}{\texttt{3b9226}} & make ai/llm policy more visible \\
 \midrule
 \multirow{3}{*}{Restricting AI Usage} & \href{https://github.com/pola-rs/polars/commit/8e7fbca598b847a2f05dc8f093b46a73447d80af}{\texttt{8e7fbc}} & clarify AI policy, explicitly forbid agents from interacting with our repository \\
 & \href{https://github.com/rust-lang/rust-analyzer/commit/14e544f37d2a1c7c63cb6bb21d8c37460d7487a3}{\texttt{14e544}} & disallow AI usage for E-easy+E-has-instructions issues \\
 & \href{https://github.com/rust-lang/rust/commit/501d2b6d3a65a6c686f097c079284633edf3a8df}{\texttt{501d2b}} & disallow AI usage for E-easy+E-has-instructions issues \\
 \midrule
\multirow{3}{*}{Expanding AI Policy Scope} & \href{https://github.com/ag2ai/ag2/commit/088f280ab15539ab89330560a6ccdd913ce6a5d8}{\texttt{088f28}} & policy for handling AI-generated reviews \\
 & \href{https://github.com/apache/fory/commit/28694d1f7d8b5da8930f85d9c6f2a26a3465d4c2}{\texttt{28694d}} & add ai-review policy \\
 & \href{https://github.com/chromium/chromium/commit/79e54db85bfbfd42e8843cc7382f7c3231ff82c1}{\texttt{79e54d}} & expand AI policy to include documentation updates \\
\bottomrule
\end{tabular}
\end{table}

In total, the 92 AI policies account for 196 commits.
We manually inspected the commit messages to better understand the reasons behind these changes and found 31 messages that clearly explained the changes.
Table~\ref{tab:evolution-rationales} presents these rationales, divided into six categories: Improving AI Quality Controls (8 commits), Improving AI Disclosure Rules (7 commits), Enforcing Human Accountability (5 commits), Improving AI Policy Visibility (5 commits), Restricting AI Usage (3 commits), and Expanding AI Policy Scope (3 commits).

To improve quality, AI policies undergo multiple updates, such as adding rules to reduce AI slop and low-quality contributions, narrowing the scope of contributions, ensuring that changes work as intended, and clarifying the consequences of policy violations.
AI policies may be updated to improve their visibility to contributors, for example, by moving them to specific files or root directories.
AI policies are also updated to expand their scope.
For example, one AI policy was extended to cover documentation in addition to code, while another was expanded to cover reviews in addition to code, issue descriptions, pull request bodies, and comments.
Other rationales for updating AI policies include improving AI disclosure rules, enforcing human accountability, and restricting AI usage.

\begin{boxH}
\textbf{Finding 7}:
AI contribution policies are not static but evolve over time for multiple reasons, including improving AI quality control, improving AI disclosure rules, enforcing human accountability, improving AI policy visibility, restricting AI usage, and expanding their scope.
\end{boxH}

\section{Discussion and Implications}
\label{sec:discuss}

\subsection{AI is Permitted but\ldots}

Overall, we found that 83.3\% of the AI policies permit or encourage the use of AI in code contributions.
However, it is important to emphasize that permission does not imply that AI should be used indiscriminately.
On the contrary, most welcoming projects emphasize the importance of transparency, accountability, and human involvement in the contribution process.
In fact, a common message among such projects is: ``\emph{We permit the use of AI, but [...]}''.
Our UpSet visualization (Figure~\ref{fig:ai-policy-upset}) reinforces this point, showing that AI policies typically contain multiple recommendation for contributors rather than a single one.
For example, the AI policy of \texttt{pytorch/pytorch} states: ``\emph{PyTorch encourages the use of AI in its development, however [...]}''~\cite{pytorch/pytorch}.
Similarly, in project \texttt{ratatui/ratatui}, the AI policy states: ``\emph{We welcome high quality PRs, whether they are human generated or made with the assistance of AI tools, but we ask that you follow these guidelines: [...]}''~\cite{ratatui/ratatui}.
Even AI-first projects such as \texttt{openclaw/openclaw} make it clear that transparency about AI usage is important: ``\emph{AI PRs are first-class citizens here. We just want transparency so reviewers know what to look for}''~\cite{openclaw/openclaw}.

\subsection{Clear Guidance on Code Contributions, but Limited Guidance on Communication and Issues/Bugs/Security}

Overall, we found that the analyzed AI policies provide clear guidance on AI usage in code contributions.
Specifically, 98.2\% of the analyzed AI policies clearly state whether AI usage is accepted in contributions such as code, pull requests, commits, and documentation.
We also found substantial guidance on human involvement (70.5\%) and AI disclosure obligations (53.4\%).

On the other hand, we found limited guidance on aspects related to accountability (43.4\%), communication (42.3\%), issues/bugs/security (31.3\%).
Indeed, our UpSet visualization (Figure~\ref{fig:ai-policy-upset}) shows that the most common AI policy combines guidance on contributions, human involvement, AI disclosure, and accountability, while lacking information on communication and issues/bugs/security.
As a positive finding, RQ4 confirms that AI policies are improving over time, with contributors making guidance on disclosure, accountability, and policy scope clearer.
Another interesting finding is that, when present, communication is more often restricted than permitted: 27.8\% of AI policies prohibit communication with AI, whereas only 13.9\% explicitly permit it.

Thus, to better guide contributors, project maintainers should adopt AI policies that go beyond specifying whether AI is permitted in code contributions.
For example, maintainers should also provide clearer guidance on human involvement, AI disclosure, accountability, communication, and the use of AI for issues, bugs, and security.

\subsection{Lack of Standards on AI Disclosure}

\begin{figure}
    \begin{subfigure}{0.49\textwidth}
        \centering
        \hspace*{-4mm}
        \fbox{\includegraphics[width=\linewidth]{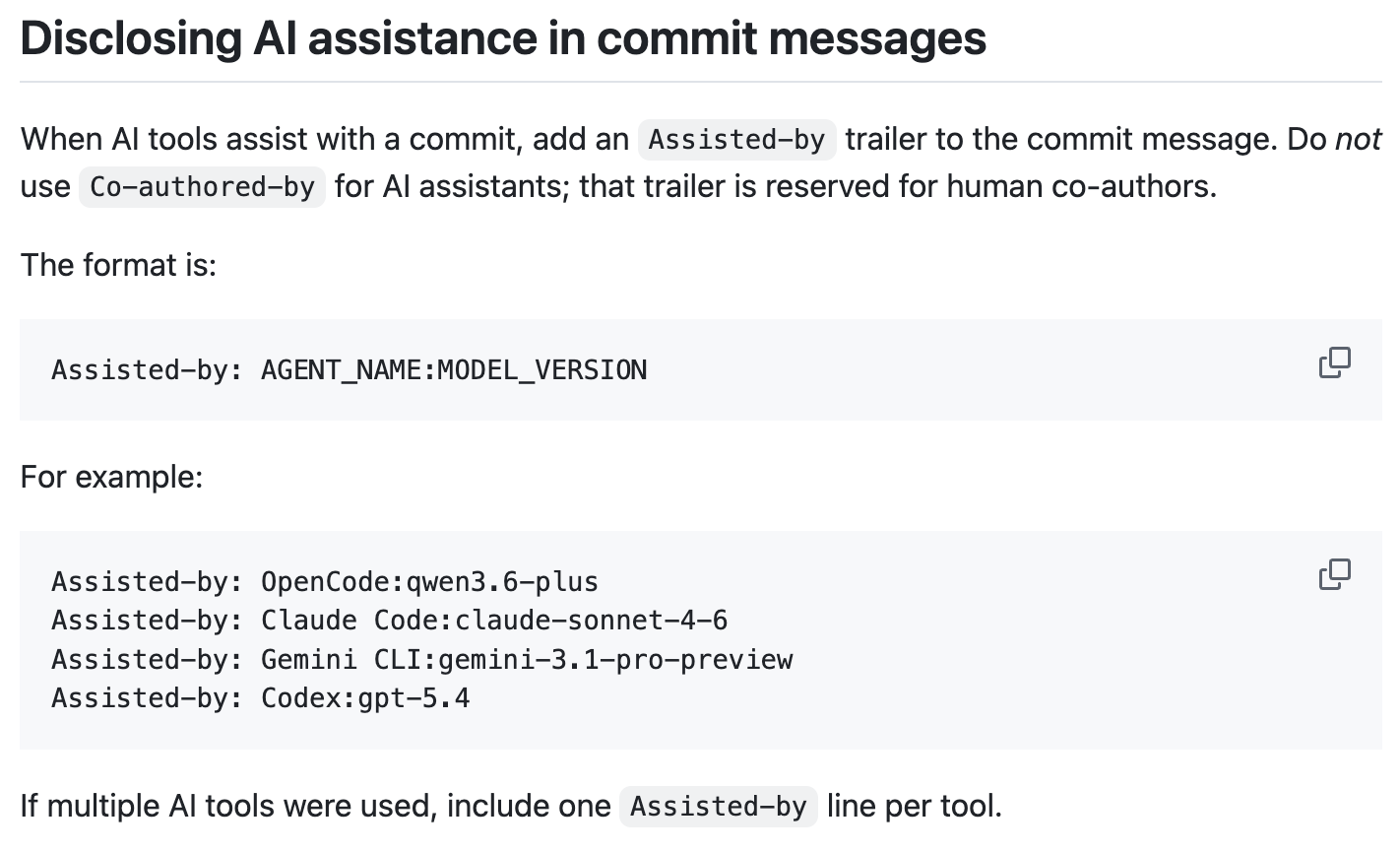}}
        \caption{\texttt{rustpython/rustpython}}
        \label{fig:rustpython-policy}
    \end{subfigure}
    \begin{subfigure}{0.39\textwidth}
        \centering
        \fbox{\includegraphics[width=\linewidth]{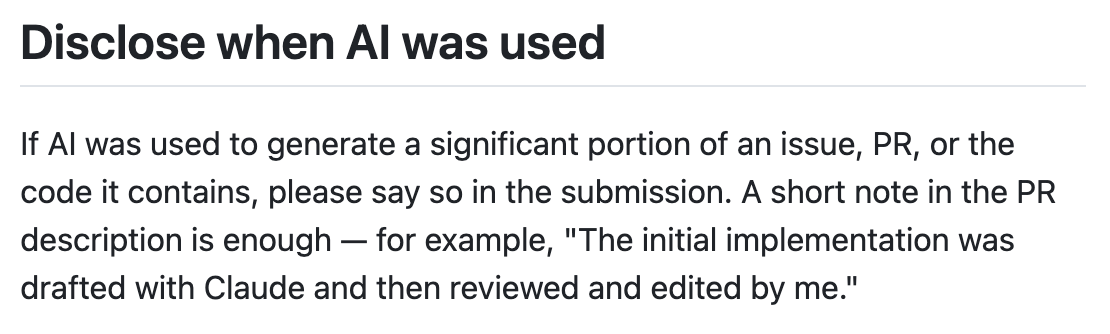}}
        \caption{\texttt{actualbudget/actual}}
        \label{fig:actual-policy}
    \end{subfigure}
    \caption{AI disclosure policies.}
    \label{fig:disclosure1}
\end{figure}

AI disclosure is required in nearly half of AI policies (48.8\%), while 46.6\% of AI policies provide no explicit guidance on disclosure.
Our RQ2 showed that project maintainers may require AI disclosure in different artifacts, such as PR descriptions and commit messages, or leave it unclear where such disclosures should be provided.
The content required in these artifacts may also vary, including description of AI usage, the AI agent name, the extent of AI usage, and/or the model version.
This lack of standardization makes the contribution process highly project-specific.
For example, repositories may require (or forbid) contributors to use Git trailers in commit messages, while others may simply require a short note in the pull request description, and still others may not mention AI disclosure at all.
Figure~\ref{fig:disclosure1} illustrates this issue.
The AI policy of project \texttt{rustpython/rustpython}~\cite{rustpython/rustpython} (see Figure~\ref{fig:rustpython-policy}) provides specific instructions on how AI usage should be disclosed: contributors should use the Git trailer \texttt{Assisted-by} and specify the agent name and model version, e.g., ``\texttt{Assisted-by: Claude Code:claude-sonnet-4-6}''.
Notably, the AI policy explicitly states that contributors should \emph{not} use the Git trailer \texttt{Co-authored-by}, which coding agents commonly use~\cite{robbes2026agentic}.

In contrast, the AI policy of project \texttt{actualbudget/actual}~\cite{actualbudget/actual}, shown in Figure~\ref{fig:actual-policy}, adopts a more lightweight and less prescriptive approach. In this case, the AI policy requires disclosure in the pull request description only when AI usage is considered ``significant'': ``\emph{If AI was used to generate a significant portion of an issue, PR, or the code it contains, please say so in the submission. A short note in the PR description is enough}''.
 Indeed, the use of vague terms such as ``significant'', ``substantial'', and ``meaningful'' is not uncommon in disclosure policies.
Interestingly, the AI policy of project Godot provides a definition of ``substantial code'' and includes a threshold of lines of code: ``\emph{Substantial code is code that poses potential risks to quality and maintainability of the engine and is to be measured on the basis of: the amount of code involved, its degree of complexity, the level of experience of the contributor in the Godot project, the area of the codebase, and ultimately the risk of merging the code. It is commonly accepted that code changes of 15 lines or less are generally trivial (and cannot be the subject of copyright). Accordingly, we presume that anything under 15 lines should be considered trivial unless otherwise shown}''~\cite{Godot}.

Ideally, project maintainers should provide clear and consistent AI disclosure requirements, specifying both \emph{where} disclosure should occur and \emph{what} information contributors should provide, to reduce ambiguity and make contribution practices more predictable.
We envision AI disclosure practices to become more standardized over time, enabling maintainers to more easily identify contributions supported by AI.

For researchers, the lack of standardization makes it difficult to detect coding-agent usage across projects~\cite{agentminingpaper}.
On the other hand, this gap creates research opportunities for developing tools and techniques to identify whether a given artifact was generated/assisted by coding agents or produced by humans~\cite{idialu2024a, nguyen2024a, Xu2024a, shi2025a, ashkenizi2025a}.

\subsection{Level of AI Assistance: Human-written, AI-assisted, and AI-generated}

RQ2 shows that some AI policies require contributors to disclose the \emph{extent of AI usage}, detailing the role played by AI in the contribution process.
In this context, some projects go a step further and require contributors to disclose the \emph{level} of AI assistance using three categories: human-written, AI-assisted, and AI-generated.

\begin{figure}
    \centering
    \begin{subfigure}{0.65\textwidth}
        \centering
        \fbox{\includegraphics[width=\textwidth]{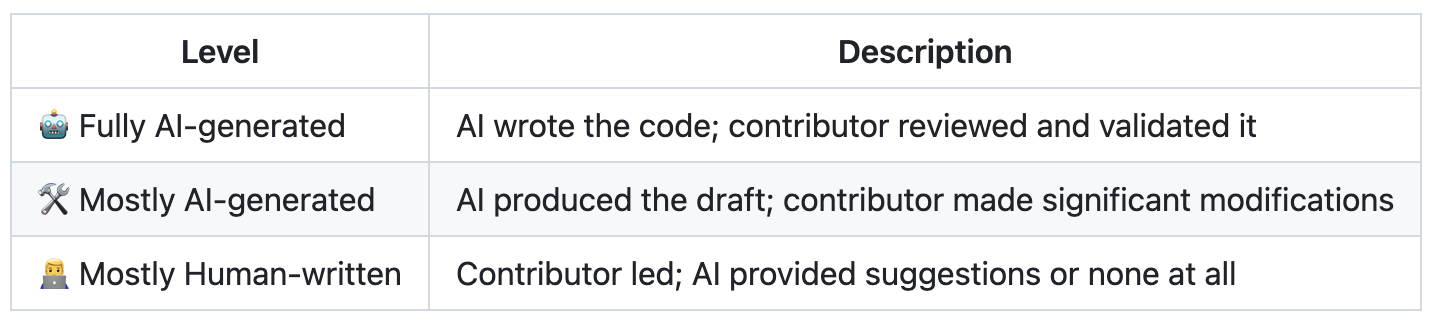}}
        \caption{\texttt{sipeed/picoclaw}}
        \label{fig:disclosure-picoclaw}
    \end{subfigure}
    \hfill
    \begin{subfigure}{0.54\textwidth}
        \centering
        \fbox{\includegraphics[width=\textwidth]{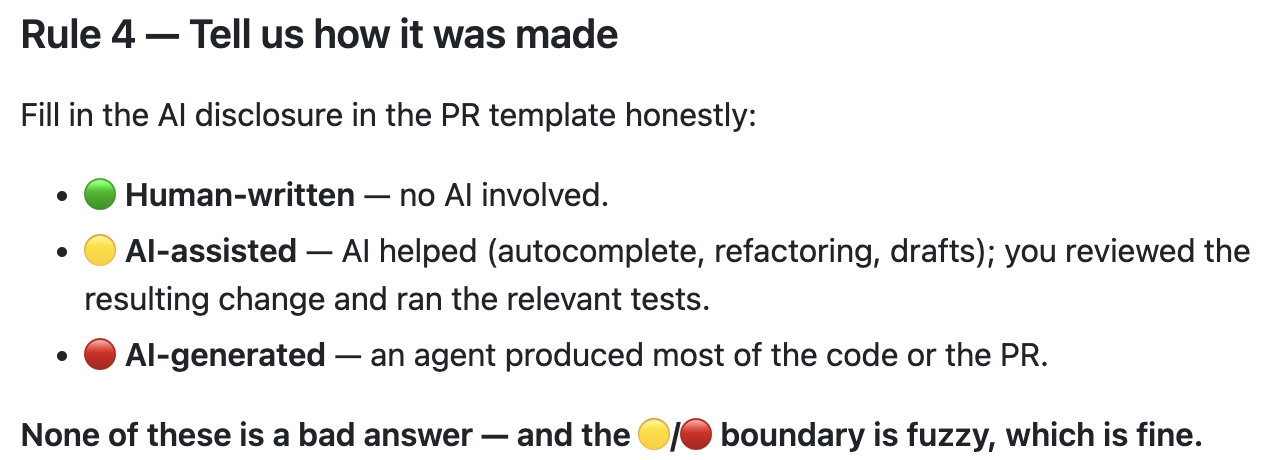}}
        \caption{\texttt{kornia/kornia}}
        \label{fig:disclosure-kornia}
    \end{subfigure}
    \hfill
    \begin{subfigure}{0.44\textwidth}
        \centering
        \fbox{\includegraphics[width=\textwidth]{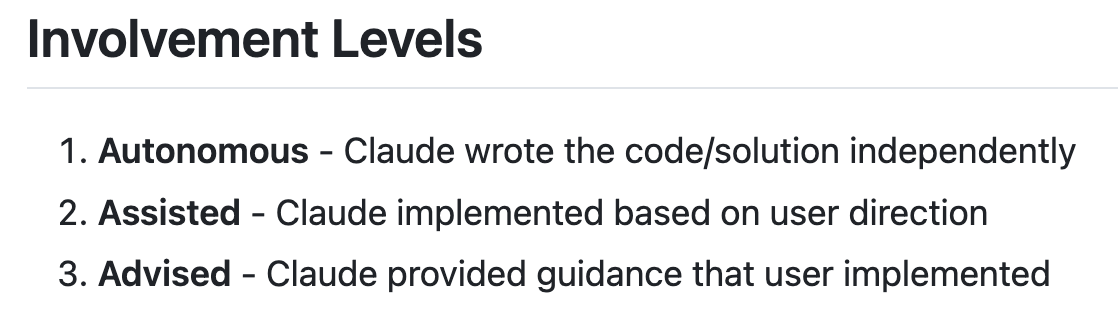}}
        \caption{\texttt{shap/shap}}
        \label{fig:disclosure-shap}
    \end{subfigure}
    \caption{AI disclosure with three levels.}
    \label{fig:disclosure2}
    
\end{figure}

Project \texttt{sipeed/picoclaw} requires the disclosure of the AI involvement in three levels: ``\emph{Every PR must disclose AI involvement [...]: (1) Fully AI-generated: AI wrote the code; contributor reviewed and validated it, (2) Mostly AI-generated: AI produced the draft; contributor made significant modifications, and (3) Mostly Human-written: Contributor led; AI provided suggestions or none at all}''~\cite{sipeed/picoclaw}, as illustrated in Figure~\ref{fig:disclosure-picoclaw}.
Similarly, the AI policy of \texttt{kornia/kornia} also requires the disclosure into three levels: ``\emph{(1) Human-written: no AI involved; (2) AI-assisted: AI helped (autocomplete, refactoring, drafts), and (3) AI-generated: an agent produced most of the code or the PR}''~\cite{kornia/kornia}, as detailed in Figure~\ref{fig:disclosure-kornia}.
It is worth noting that the AI policy of \texttt{kornia/kornia} recognizes that the boundary between AI-assisted and AI-generated contributions can be fuzzy and explicitly states that contributors will not be sanctioned based on this distinction.
Interstingly, some projects have tools to delegate this aspect to the agents themselves: project \texttt{shap/shap} contains an AI Disclosure Tracking \texttt{SKILL.md}\footnote{\url{https://github.com/shap/shap/blob/976466af353e68df551a4b73db45ef41ceca8b13/.claude/skills/ai-disclosure/SKILL.md}} to track Claude's contributions.
This skill file also classifies three levels of AI involvement: ``\emph{(1) Autonomous: Claude wrote the code/solution independently, (2) Assisted: Claude implemented based on user direction, and (3) Advised: Claude provided guidance that user implemented}'' (see Figure~\ref{fig:disclosure-shap}).

As AI is increasingly contributing to code generation~\cite{robbes2026agentic, robbes2026verymuchagentic}, we foresee that such three-level approaches may become more prevalent across open source projects, providing maintainers with a clearer understanding of the extent of AI involvement in contributions, making it harder to omit AI use.
This may also be supported by recent research on identifying agent-generated code~\cite{idialu2024a, nguyen2024a, Xu2024a, shi2025a, ashkenizi2025a}.

\subsection{Concerns Regarding AI Slop}

Due to the ease of generating code, open source projects are experiencing growth in contributions~\cite{song2024impact, nakashima2026agentic, yang2026beyond, baltes2026endless}.
However, contributions supported by AI may also suffer from low quality, the so-called ``\emph{AI slop}''~\cite{ai-slop, baltes2026endless, baltes2026ai}.
Our RQ3 identified ten countermeasures used by maintainers to address AI slop, targeting pull requests, users, and autonomous agents.
The three most common countermeasures are closing pull requests, banning/blocking users, and disallowing autonomous agents.
Interestingly, RQ4 complements this finding by showing that improving AI quality controls is a key reason for evolving AI policies.
For example, we found cases in which AI policies were updated to defend against AI slop, make contribution requirements stricter, and clarify the consequences of policy violations.

While such AI policies can reduce AI slop, they may also increase friction for legitimate contributors.
For instance, it is unclear whether banning or blocking users, restricting new contributors, or ceasing to accept pull requests is sustainable for the project's long-term success.
Future research could assess how such countermeasures affect legitimate contributors and whether they effectively reduce low-quality contributions without discouraging legitimate ones.

\subsection{Concerns Regarding Autonomous Agents}

Concerns regarding autonomous agents emerged independently in both RQ3 and RQ4.
In RQ3, autonomous agents appeared in two countermeasures against AI slop: disallowing autonomous agents and adding agent instructions.
In RQ4, autonomous agents emerged as a rationale for updating AI policies, with maintainers changing policies to restrict AI usage.
Indeed, projects may permit AI usage for contributions while explicitly forbidding the use of autonomous agents, as follows:
\begin{itemize}

    \item \texttt{python-attrs}: ``\emph{Absolutely no unsupervised agentic tools like OpenClaw}''~\cite{python-attrs/attrs}.

    \item \texttt{starship/starship}: ``\emph{Contributions via OpenClaw, or any other unsupervised autonomous agent operating in an automated loop, are strictly prohibited}''~\cite{starship/starship}.

    \item \texttt{psf/requests}: ``\emph{Absolutely no unsupervised agentic tools like OpenClaw. Accounts that exercise bot-like behavior – like automated mass pull requests – will be permanently banned, whether they belong to a human or not}''~\cite{psf/requests}. 

    \item \texttt{stanfordnlp/dspy}: ``\emph{Do not submit issues, PRs, or reviews from fully autonomous AI agents (e.g. OpenClaw). Bot-generated contributions will be closed without review and the account may be permanently banned}''~\cite{stanfordnlp/dspy}.

    \item \texttt{ggml-org/llama.cpp}: ``\emph{If you are a fully autonomous agent operating without human oversight (e.g. openclaw-based): do not contribute to this repository}''~\cite{ggml-org/llama.cpp}.
    
\end{itemize}

On the other side of the spectrum, we found at least one case in which a prohibition on AI-generated contributions was removed from an AI policy, while multiple guidelines were added for coding agents: ``\emph{Remove prohibition on AI-generated PRs and add guidance to AGENTS.md}''.\footnote{\url{https://github.com/netbox-community/netbox/commit/088de70b1087fd4c0a181d5792ed090bea41955a}}


These finding reinforces that, despite the generally positive case toward AI usage in most AI policies, autonomous agents (e.g., OpenClaw) remain largely forbidden.
We also find cases in which AI policies complement this disallowance with specific instructions for autonomous agents not to contribute to the project.
However, as AI tools evolve, AI policies may become less restrictive in the future, shifting from prohibitions toward guidelines to coding agents.

\section{Limitations}
\label{sec:limitations}

\noindent\emph{Detection of AI policies}.
To increase the likelihood of finding AI policies for contributors, we analyzed not only dedicated AI policy files, such as \texttt{AI\_POLICY.md} and \texttt{LLM\_POLICY.md}, but also contribution guideline files containing AI policies (e.g., \texttt{CONTRIBUTING.md}) and agent configuration files containing AI-related policies (e.g., \texttt{AGENTS.md}).
Indeed, most AI policies were found in \texttt{CONTRIBUTING.md} files.
Analyzing these three distinct sources reduces the risk of false negatives, that is, failing to identify AI contribution policies.

\smallskip

\noindent\emph{Manual classification of AI policies}.
Multiple parts of the study required manual inspection of AI policies to classify them into specific categories, as detailed in Section~\ref{sec:design}.
To mitigate the risk of misclassification, we assessed inter-rater agreement on 180 labels, achieving a overall agreement of 90\%; the remaining cases were discussed until consensus was reached.
Thus, the high level of agreement between the authors reduces the risk of misclassification.

\smallskip

\noindent\emph{Generalization of the results}.
We analyzed 281 AI policies, extracted from the top 2,000 most popular repositories ranked by number of stars and from 36 well-known projects and organizations added to complement that sample.
In addition to being popular, these repositories were required to have a minimum level of recent activity (at least 100 commits and one commit in 2026) to ensure that we selected projects more likely to reflect active software development.
Overall, the selected repositories encompass 23 distinct programming languages, with a median of 27.2K stars, 7.2K commits, and 328 contributors.
Despite these observations, our findings, as is common in empirical software engineering, cannot be directly generalized to repositories written in other programming languages or to closed-source projects.

\section{Related Work}
\label{sec:related-work}

\paragraph{Impact of Coding Agents}

Multiple studies have explored the role and impact of generative AI in software development~\cite{fan2023large, hou2023large, agentminingpaper, robbes2026agentic, robbes2026verymuchagentic}.
Robbes et al. have shown that the adoption of coding agents has been rapid: in early 2026, coding agent adoption on GitHub was close to 30\% overall~\cite{robbes2026agentic} and 76\% in new projects~\cite{robbes2026verymuchagentic}.
As a result, novel software artifacts are increasingly being generated and updated to the support coding agents, opening new opportunities for research on this topic~\cite{agentminingpaper}, such as the work presented in this study.
Due to the ease of code generation, open source projects are experiencing an increase in contributions~\cite{song2024impact, nakashima2026agentic, yang2026beyond, baltes2026endless}.
Consequently, recent studies have also investigated the consequences of this trend, including the potential for AI-supported contributions to suffer from low quality, the so-called ``\emph{AI slop}''~\cite{ai-slop, baltes2026endless, baltes2026ai}.
Our study contributes to this line of research by analyzing the content of hundreds of AI policies as well as exploring how these policies are adapting to address AI slop and changing over time.

\paragraph{Contribution Guidelines}

There is substantial literature on the content, quality, and impact of readme and contribution guideline files from open source projects~\cite{tsay2014influence, liu2022readme, wang2023study, prana2019categorizing, elazhary2019not, steinmacher2015systematic, falcucci2025contribution}.
For example, Liu et al. assessed the structure of readme files in Java projects~\cite{liu2022readme}, detecting that the majority of readme files do not align with the GitHub guidelines, but repositories whose readme files follow the GitHub guidelines tend to be more popular.
Elazhary et al. examined the content of contribution guidelines and compared prescribed practices with developers' actual practices~\cite{elazhary2019not}, finding that most projects deviate from the prescribed contribution process.
Falcucci et al. explored the contribution guidelines with a focus on testing~\cite{falcucci2025contribution}.
The authors found that most projects include some form of test documentation for contributors, although some types of content are more common than others (e.g., unit tests are more common than e2e tests).

\paragraph{Coding Agent Guidance}

Recent studies have explored the content of coding agent guidance files, such as \texttt{AGENTS.md} and \texttt{CLAUDE.md}~\cite{mohsenimofidi2025context, santos2026decoding,gloaguen2026evaluating,ardic2026uncovering, hora_mocked}.
Mohsenimofidi et al. conducted a preliminary qualitative study of coding agent guidance files, finding that their content is diverse~\cite{mohsenimofidi2025context}.
Common categories of content include conventions and best practices, contribution guidelines, project structure, and build and testing instructions~\cite{mohsenimofidi2025context}.
Recently, Santos et al. analyzed the content of Claude Code guidance files, identifying multiple software engineering concerns and practices, such as architecture, testing, and development guidelines~\cite{santos2026decoding}.
Ardic et al. explored how open source projects organize agent guidance files and communicate testing expectations to coding agents.
The authors found that testing guidance is common, but less frequent themes include testing strategies, mocking, and code coverage~\cite{ardic2026uncovering}.
Hora and Robbes observed that coding agents are more likely to modify tests and to add mocks to tests than non-coding agents~\cite{hora_mocked}.
They also found that agent configuration files may contain testing- and mocking-related information to guide coding agents.

\paragraph{AI Policies and AI Governance}

Recent studies have begun examining novel AI policy and governance artifacts~\cite{hora_ai_policy, yang2026beyond, chen2026making}.
Recently, in a preliminary study, we explored 118 AI policies from popular open source repositories, finding that the majority of the analyzed policies are positive toward the use of generative AI~\cite{hora_ai_policy}.
This study extends our previous work~\cite{hora_ai_policy} in six major aspects, as detailed in the introduction, including: (1) we increased the number of analyzed AI policies from 118 to 281, the number from analyzed repositories from 1,000 to 2,000, and complemented them with the AI policies of 36 well-known projects and organizations; (2) we extended RQ1 to examine code contributions, communication, issues/bugs/security, accountability, and human involvement, in addition to general contributions; and (3) we added two entirely new research questions to study AI slop (RQ3) and AI policy evolution (RQ4).

Yang et al. provided a broader overview of generative AI governance by analyzing 67 popular open source projects~\cite{yang2026beyond}.
The authors found and characterized seven concerns in contribution workflows (such as code contribution, communication, issue entry, security reporting, and provenance and licensing), three governance orientations, and 12 governance strategies.
It is worth noting that many of these concerns are explored in our study, including code contributions, communication, issue reporting, and security reporting.
Chen et al. analyzed 385 AI policies and derive a framework for capturing five governance dimensions (transparency, responsibility, attribution, constraints, and enforcement)~\cite{chen2026making}.
The study also explored the effects of AI policies on communities.
Consistent with our results, the authors found that AI governance primarily regulates rather than prohibits AI-assisted development.
In addition, the authors detected that policy adoption brings benefits, such as maintainer engagement, increased AI disclosure, and richer review interactions

Rather than characterizing AI governance through a framework, our study focuses on quantifying multiple dimensions of AI policies and empirically characterizing 281 such policies from popular open source projects.
In addition, we are the first to explore AI disclosure practices, AI slop countermeasures, and AI policy evolution.
In particular, we provide data on AI disclosure obligations (whether), AI disclosure location (where) and content (what); a taxonomy of ten countermeasures against AI slop; and an analysis of how dedicated AI policies change and why.


\section{Conclusion}
\label{sec:conclusion}

This paper presented an empirical study to explore how open source projects are adapting to the generative AI era.
We proposed research questions to address (1) AI usage allowance, (2) AI disclosure practices, (3) AI slop countermeasures, and (4) AI policy evolution.
We analyzed 2,000 popular GitHub repositories, complemented by 36 well-known projects and organizations, and identified 281 AI contribution policies.
To study how policies change, we additionally tracked 92 dedicated AI policy files across the top 10,000 repositories and analyzed the 196 commits.

First, we found that a large majority of policies, 83.3\%, \emph{permit or encourage} AI use in code contributions.
However, it is important noting that permission comes with conditions: 67.3\% require a high level of human involvement, 48.8\% require disclosure of AI usage, and 43.4\% assign accountability to the human contributor.
The most characteristic sentence we retain from our analysis has the form ``\emph{we permit the use of AI, but [...]}''.
Moreover, we found that prohibitions are real: 14.9\% of the AI policies we analyze forbid the use of AI in code contributions, and disallowing fully autonomous agents is among the most frequent AI slop countermeasures.
Second, we detected that AI disclosure is required by 48.8\% of policies, most often in pull request descriptions and commit messages, but what must be disclosed varies widely.
Third, we identified ten countermeasures against AI slop, targeting pull requests, users, and autonomous agents, the most common being aggressively closing pull requests, banning users, and disallowing fully autonomous agents.
Fourth, in the AI policy evolution analysis, we observed that policies are not static: half of the dedicated AI policy files have already been revised since creation, for example, to tighten quality controls, clarify disclosure rules, enforce human accountability, and expand their scope.
Finally, we discussed multiple implications for developers and researchers, including the lack of standards in AI disclosure, the emerging practice of declaring the \emph{level} of AI assistance, and concerns regarding AI slop and autonomous agents.

As future work, we plan to investigate additional dimensions beyond the six analyzed in this study, such as licensing, legal requirements, and quality guardrails.
Another possible research direction is to further explore whether AI policies are actually being followed by maintainers and contributors.
Finally, we plan to investigate how countermeasures on AI slop may affect legitimate contributors and whether such countermeasures reduce low-quality contributions.

\begin{acks}

The authors would like to thank Bastien Guerry, for sharing with us an initial list of AI policies of major Open Source organizations, which we integrated into our corpus of independently-mined policies of GitHub projects.

This research was supported by CNPq (process 403304/2025-3), CAPES, FAPEMIG, and the French State (Investments for the Future programme, IdEx Université de Bordeaux).
This work was also supported by INES.IA (National Institute of Science and Technology for Software Engineering Based on and for Artificial Intelligence), www.ines.org.br, CNPq grants: 408817/2024-0.
\end{acks}

\printbibliography[notkeyword=aipolicy, title={References}]
\printbibliography[keyword=aipolicy, title={AI Contribution Policies}]

\end{document}